\documentclass[prd,aps,a4paper,twocolumn,eqsecnum,nofootinbib,floatfix]{revtex4} 

\newif\ifusesec
\usesectrue  

\usepackage{amsmath,amsfonts,amssymb}

\newcommand{\beq}{\begin{equation}}
\newcommand{\eeq}{\end{equation}}
\newcommand{\bea}{\begin{eqnarray}}
\newcommand{\eea}{\end{eqnarray}}
\begin{document}

\title{Radiation-reaction driven dynamics at the third-and-a-half post-Newtonian order in different gauges}
\author{Donato Bini$^{1}$, Andrea Geralico$^{1}$, Sara Rufrano Aliberti$^{2,3}$}  
  \affiliation{
$^1$Istituto per le Applicazioni del Calcolo ``M. Picone,'' CNR, I-00185 Rome, Italy\\
$^2$Scuola Superiore Meridionale, Largo San Marcellino 10, 80138, Naples, Italy\\
$^3$INFN, Sezione di Napoli, Complesso Universitario di Monte S. Angelo, Via Cinthia Edificio 6, 80126, Naples, Italy
}

\date{\today}

\begin{abstract}
We consider the coordinate-dependent definition of the radiation-reaction force at the third-and-a-half post-Newtonian order for general orbits. In order to (partially) determine its expression we refer to the balance method, involving the energy and angular momentum lost by the system, the Schott terms, and the energy and angular momentum fluxes at infinity.
Only the latter are gauge-invariant quantities when passing from a coordinate system to another. 
The gauge dependence of both radiation-reaction force and Schott terms is encoded in a set of gauge parameters entering their definitions.
We show how to relate the harmonic-coordinate losses of mechanical energy and angular momentum by the system with the radial and azimuthal components of the radiation-reaction force in a different coordinate system expressed in terms of phase-space variables in a Hamiltonian framework.
The advantage of this approach is that only the coordinate transformation between harmonic coordinates and coordinates and momenta in the new coordinate system is needed, without solving again the balance equations.
We derive such a transformation for both Arnowitt-Deser-Misner  and Effective-One-Body  coordinates.
In the latter case we also discuss some simplifying choices of the gauge parameters adopted in current waveform models.
Finally, we show how to obtain the solution for the radiation-reaction correction to the orbit in the new coordinate system simply by transforming the harmonic-coordinate solution known in the literature. 
This is a remarkable simplification, since one can avoid solving again for the radiation-reacted dynamics.
\end{abstract}

\maketitle

\section{Introduction}

The derivation of the equations of motion and their solution constitutes the primary task in the study of the two-body dynamics
as it is one of the building blocks for any waveform model.
Working within a post-Newtonian (PN) framework and in the center-of-mass (cm)  frame the relative motion is conservative up to the 2PN accuracy, whereas starting from the 2.5PN order one must take into account the presence of a radiation-reaction (rr)   force. It is well known that incorporating dissipative effects makes the analysis of the dynamics considerably more involved.
Up to the 3.5PN order the contribution of the rr force is purely instantaneous. Nonlocal (hereditary) effects (tail terms) first appear at the 4PN order (see, e.g., Ref. \cite{Blanchet:2013haa} for a review).
The leading order (LO) rr force was first obtained in harmonic coordinates in Refs. \cite{Damour:1981bh,D1982,Damour:1983tz} directly solving the two-body dynamics by integrating the retarded field generated by the source.
This first-principle approach has been later extended to the next-to-leading order (NLO) in Ref. \cite{Jaranowski:1996nv} by using Arnowitt-Deser-Misner (ADM) coordinates.
NLO results in harmonic coordinates have also been obtained in Refs. \cite{Pati:2002ux,Nissanke:2004er} by a matching procedure between the near-zone and the wave-zone field.
The current knowledge of the rr force is the 4.5PN order, including nonlocal effects due to the recoil of the source \cite{Blanchet:1996vx,Blanchet:2024loi,Blanchet:2026suq}.
So far, independent calculations were done within the effective field theory (EFT) formalism at the same PN accuracy (see Refs. \cite{Galley:2012qs,Leibovich:2023xpg}).

A simpler, more direct method to (partially) determine the rr force has been proposed by Iyer and Will (IW) \cite{Iyer:1995rn} by assuming a balance equation between the (gauge-invariant) instantaneous energy and angular
momentum fluxes in the far zone and the instantaneous (coordinate-dependent) energy and angular momentum losses in the system's near zone. 
This procedure finally leads to an expression for the  rr  force depending on a set of gauge parameters, which cannot be determined unless one is able to derive the equations of motion of the binary from first principles, i.e., by directly integrating the field equations in a given coordinate system (as it actually happened in the cases of harmonic, ADM and Burke-Thorne (BT) coordinates). The roots of this ambiguity should be traced back into the balance laws themselves, which relate gauge-invariant fluxes at infinity of energy and angular momentum with the sum of the energy and angular momentum included both in the system and in the radiation field (Schott terms), only this sum having a gauge invariant meaning and not the two contributions separately (with the possibility to move terms from one contribution to the other).

Iyer and Will also discussed how to construct the rr force by using the 3.5PN near-zone rr potentials valid in an extended BT gauge obtained by Blanchet \cite{Blanchet:1993ng} through the multipolar-post-Minkowskian formalism, once expressed in terms of time derivatives of the source multipole moments. 
Comparison of the resulting  rr  force in the cm frame with the general one obtained by the balance method allows for fixing the values of the unknown parameters corresponding to the BT gauge.
The harmonic coordinate values of the gauge parameters at the 3.5PN order (extending previous results by Damour and Deruelle \cite{Damour:1981bh,D1982,Damour:1983tz} at the leading 2.5PN order) have been determined in Refs. \cite{Pati:2002ux,Nissanke:2004er} through the PN iteration of the Einstein field equations complemented by the harmonic coordinate condition by using two different implementations of the asymptotic matching procedure, but leading to the same result.
The ADM coordinate values have instead been obtained in Ref. \cite{Konigsdorffer:2003ue}, where the 3.5PN binary's equations of motion and the corresponding rr force have been derived within the framework of the ADM Hamiltonian formalism.
The values of the gauge parameters corresponding to these coordinate systems are listed in Table \ref{tab1:deltas}.

\begin{table*}
\caption{\label{tab1:deltas} Different coordinate choice and gauge parameters at the 3.5PN level.
The latter are denoted as $\alpha,\beta$ at 2.5PN, and $\delta_i$, $i=1\ldots 5$, and $\epsilon_5$ at 3.5PN, following the notation of Ref. \cite{Iyer:1995rn}.
}
\begin{ruledtabular}
\begin{tabular}{lll|llllll}
Coordinates &$\alpha$ &$\beta$& $\delta_1$& $\delta_2$& $\delta_3$& $\delta_4$& $\delta_5$& $\epsilon_5$\\
\hline
h& $-1$ &$0$& $\frac{271}{28}+6\nu$& $-\frac{77}{4}-\frac32 \nu$& $\frac{79}{14}-\frac{92}{7}\nu$& $10$& $\frac{5}{42}+\frac{242}{21}\nu$& $-\frac{439}{28}+\frac{18}{7}\nu$\\
BT& $4$ &$5$& $-\frac{99}{14}+27\nu$& $5(1-4\nu)$& $\frac{274}{7}+\frac{67}{21}\nu$& $\frac52 (1-\nu)$& $-\frac{292}{7}-\frac{57}{7}\nu$& $\frac{51}{28}+\frac{71}{14}\nu$\\
ADM  & $\frac53$ &$3$& $\frac{41}{84}+\frac{677}{42}\nu$& $-\frac{61}{14}-\frac{151}{14}\nu$& $\frac{583}{28}-\frac{157}{21}\nu$& $\frac{115}{28}-\frac{15}{7}\nu$& $-\frac{961}{42}+\frac{16}{3}\nu$& $-\frac{51}{14}-\frac{23}{14}\nu$\\
\end{tabular}
\end{ruledtabular}
\end{table*}

For the various coordinate systems mentioned above (i.e., harmonic, ADM and BT coordinates), the rr force is completely determined by solving the equations of motion of each body, expressing then the relative acceleration in the cm frame.  
This first-principle approach then proves the validity of the balance laws at a given PN order.
This is enough to fix all gauge parameters entering the IW general expression for the rr force without any residual freedom. 
Using instead only the balance approach there is no way to uniquely determine the gauge parameters.
This is the case of the Effective-One-Body (EOB) formalism \cite{Buonanno:1998gg,Buonanno:2000ef}, which only exists in the cm frame.
The gauge parameters can then be chosen arbitrarily, and various choices associated with different motivations are available in the literature, with the aim of either reducing the freedom in the definition of Schott terms, or leading to simplified expressions for the various quantities. Clearly,  whatever is the choice performed, it necessarily affects the dynamics (without modifying any gauge-invariant quantity), since the corresponding description uses coordinates which are gauge-dependent objects.

Bini and Damour (BD) \cite{Bini:2012ji} formulated a balance approach by using a Hamiltonian framework with the aim of translating into the EOB formalism the  rr  force derived by Iyer and Will and later extended to the 4.5PN order in Ref. \cite{Gopakumar:1997ng} (still neglecting nonlocal effects). BD imposed the condition that the Schott contribution to the angular momentum vanishes identically, which completely determines the azimuthal component of the rr force, but only corresponds to part of the freedom found by Iyer and Will. The balance equations then fix the radial component of the rr force as well as the Schott contribution to the energy, with a residual gauge freedom, which can be used to get simplified expressions.
Following the BD approach other gauge choices have been proposed still within the EOB framework \cite{Khalil:2021txt,Ramos-Buades:2021adz,Gamboa:2024imd}. We will shortly review these choices too. 

The aim of the present work is to provide a general prescription to translate into a Hamiltonian framework the final results for the rr force and Schott terms obtained in harmonic coordinates by following the IW approach.
To this end it is enough to identify the force components in the new formulation in terms of the instantaneous variations of the energy and angular momentum of the system due to radiation reaction in the old one, applying then the coordinate transformation between harmonic coordinates and coordinates and momenta of the new Hamiltonian coordinate system, without solving again the balance equations as in the BD approach. 
Our results will be also left with the IW gauge parameters unspecified, so that whatever gauge choice can be performed at any moment in a simple way.
Furthermore, one can immediately obtain the solution for the rr correction to the orbit in the new coordinate system simply by inverting the above transformation and using the already known solution in harmonic coordinates available in the literature (see Ref. \cite{Damour:2004bz} for the bound case and Ref. \cite{Bini:2025rng} for unbound orbits).
In this way one can avoid solving again the radiation-reacted dynamics.
In fact, even if radiation-reaction effects are usually treated perturbatively (conservative plus radiation-reaction) by using the Lagrange method of variation of arbitrary constants, one should have to solve a system of coupled differential equations for the variation of the orbital parameters in any chosen coordinate system, which is a hard task for increasing PN accuracy.

We will mainly work in units of $c=1=G$ and denote as $\eta=\frac{1}{c}$ a place-holder for PN expansion when needed.

\section{CM relative 3.5PN dynamics}

Let us consider a binary system made of two nonspinning compact objects with masses $m_1$ and $m_2$ moving along general orbits.
Standard notations are: total mass $M=m_1+m_2$, reduced mass $\mu=m_1m_2/M$, symmetric mass ratio $\nu=\mu/M$.

The 3.5PN relative dynamics is summarized by the relative acceleration of the two bodies, ${\mathbf a}={\mathbf a}_1-{\mathbf a}_2$
\bea
{\mathbf a}&=& {\mathbf a}_{\rm N}+{\mathbf a}_{\rm 1PN}+{\mathbf a}_{\rm 2PN}+{\mathbf a}_{\rm 2.5PN}\nonumber\\
&+&{\mathbf a}_{\rm 3PN}+{\mathbf a}_{\rm 3.5PN}+O\left(\frac{1}{c^8}\right)\,,
\eea
standardly related to the relative position vector $x^i=x_1^i-x_2^i$,  $a^i\equiv  \ddot x^i $ (a dot denoting time derivative).
The total (relative) acceleration can be split in a conservative part and a rr part, 
\bea
\label{a_cons_and_rr}
{\mathbf a}&=& {\mathbf a}_{\rm cons}+{\mathbf a}_{\rm rr}\,,
\eea
with $\boldsymbol{\mathcal F}_{\rm rr}=\mu {\mathbf a}_{\rm rr}$ the associated rr force.
Through the 3.5PN order we then have ${\mathbf a}_{\rm cons}= {\mathbf a}_{\rm N}+{\mathbf a}_{\rm 1PN}+{\mathbf a}_{\rm 2PN}+{\mathbf a}_{\rm 3PN}$ and ${\mathbf a}_{\rm rr}={\mathbf a}_{\rm 2.5PN}+{\mathbf a}_{\rm 3.5PN}$.

Both $a^i_{\rm cons}$ and $a^i_{\rm rr}$ are expressed in terms of vectors in the orbital plane taken as the $x$-$y$ plane, conventionally chosen as ${\mathbf n}={\mathbf x}/r$ (unit vector along the radial direction) and ${\mathbf v}$ (relative velocity). This implies that the angular momentum vector is directed orthogonally to the orbital plane, i.e., along the $z$ axis: ${\mathbf J}=J_z e_{z}\equiv J e_z$. 

One can equivalently use a Hamiltonian description of the system $H=H(x^i,p_i)$, with $(x^i,p_i)$ a set of phase-space variables.
Hamilton's equations imply
\beq
\label{eq_of_mot0}
\dot x^i =\frac{\partial H(x,p)}{\partial p_i}\,,\qquad \dot p_i=-\frac{\partial H(x,p)}{\partial x^i}+{\mathcal F}_i\,,
\eeq
where the Hamiltonian $H$ accounts for all conservative effects, while ${\mathcal F}={\mathcal F}_{\rm rr}$ represents the external (non-conservative) radiation-reaction force.
For a review of the rr force within a Lagrangian formalism see, e.g., Ref. \cite{Buonanno:2000ef}, Eq. (3.1).

\subsection{Conservative sector}

The standard form of the conservative acceleration in a given coordinate system is the following
\bea
\label{aicons}
a^i_{\rm cons}&=& -\frac{GM}{r^2}\left[{\mathcal A}_{\rm cons} n^i+{\mathcal B}_{\rm cons}\dot r v^i\right]\,,
\eea
with $\dot r={\mathbf n}\cdot{\mathbf v}$, and
\bea
{\mathcal A}_{\rm cons}&=&{\mathcal A}_{\rm N}+\eta^2 {\mathcal A}_{\rm 1PN}+\ldots
\,,\nonumber\\
{\mathcal B}_{\rm cons}&=&{\mathcal B}_{\rm N}+\eta^2 {\mathcal B}_{\rm 1PN}+\ldots\,.
\eea
The Newtonian acceleration is
\beq
{\mathbf a}_{\rm N}=-\frac{GM}{r^2}{\mathbf n}\,,
\eeq
so that ${\mathcal A}_{\rm N}=1$ and ${\mathcal B}_{\rm N}=0$.  
At the 1PN level in harmonic coordinates one has 
\bea
{\mathcal A}^{\rm h}_{\rm 1PN}&=&-2(2+\nu)\frac{GM}{r}+(1+3\nu)v^2 -\frac32 \nu \dot r^2\,, \nonumber\\
{\mathcal B}^{\rm h}_{\rm 1PN}&=&-2(2-\nu)\,,
\eea
while at the 2PN level
\bea
{\mathcal A}^{\rm h}_{\rm 2PN}&=&\frac34 (12+29\nu)\left(\frac{GM}{r}\right)^2 +\nu(3-4\nu) v^4\nonumber\\ &+&\frac{15}{8}\nu(1-3\nu)\dot r^4
-\frac32 \nu (3-4\nu)v^2\dot r^2\nonumber\\
&-&\frac12 \nu (13-4\nu)\frac{GM}{r}v^2 -(2+25\nu+2\nu^2)\frac{GM}{r}\dot r^2 
\,, \nonumber\\
{\mathcal B}^{\rm h}_{\rm 2PN}&=&\frac12(4+41\nu+8\nu^2) \frac{GM}{r} -\frac12\nu (15+4\nu)v^2\nonumber\\
&+&\frac32\nu(3+2\nu) \dot r^2\,.
\eea
At the 3PN level we will refer to modified harmonic (MH) coordinates \cite{Blanchet:2013haa}, with
\begin{widetext}
\bea
{\mathcal A}^{\rm MH}_{\rm 3PN}&=& \left( -\frac{35}{16}\nu   + \frac{175}{16} \nu^2  -  \frac{175}{16}  \nu^3 \right)  \dot r^6 
+ \left(\frac{15}{2}\nu  -\frac{135}{4} \nu^2   +  \frac{255}{8}  \nu^3 \right) \dot r^4  v^2
+\left(
  - \frac{15}{2} \nu   +  \frac{237}{8}  \nu^2  -\frac{45}{2} \ \nu^3\right) \dot r^2v^4  
 \nonumber\\ 
&+& \left(\frac{11}{4}\nu   - \frac{49}{4}\nu^2   + 13\nu^3  \right)v^6  \nonumber\\
&+& \frac{  G M}{r}\left[ \left(79 \nu -\frac{69}{2} \nu^2  - 30  \nu^3 \right)\dot r^4
+\left(
- 121 \nu + 16  \nu^2   + 20  \nu^3   \right) \dot r^2  v^2
+ \left( \frac{75}{4}\nu    
+ 8\nu^2  - 10\nu^3  \right) v^4\right] \nonumber\\
& +& \frac{ G^2M^2}{r^2}\left[\left(1 +  \frac{22717}{168}  \nu  +  \frac{11}{8} \nu^2   - 7 \nu^3 +  \frac{615}{64} \nu\pi^2\right)
\dot r^2
+\left( -  \frac{20827}{840}\nu     +  \nu^3   
- \frac{123}{64}\nu \pi^2\right) v^2 \right]\nonumber\\
&+& \frac{  G^3M^3}{r^3} \left( -16- \frac{1399}{12}\nu -\frac{71}{2}\nu^2 +\frac{41}{16}\nu\pi^2 \right) 
 \,, \nonumber\\
{\mathcal B}^{\rm MH}_{\rm 3PN}&=&  \left(-\frac{45}{8}\nu  + 15  \nu^2 + \frac{15}{4}\nu^3\right) \dot r^4  
+\left( 12  \nu   - \frac{111}{4} \nu^2    - 12 \nu^3 \right)\dot r^2 v^2  
+\left(- \frac{65}{8}  \nu     + 19  \nu^2   + 6  \nu^3\right) v^4\nonumber\\
 &+& \frac{  G M}{r}  \left[ \left(\frac{329}{6}  \nu  +  \frac{59}{2}  \nu^2   + 18 \nu^3\right)\dot r^2 +\left(- 15  \nu - 27  \nu^2  - 10  \nu^3\right) v^2\right] \nonumber\\
&+& \frac{  G^2M^2}{r^2} \left( -4  - \frac{5849}{840}  \nu    + 25  \nu^2 + 8  \nu^3-\frac{123}{32}  \nu\pi^2\right)
 \,.
\eea
\end{widetext}

\subsection{Radiative sector}

The 3.5PN rr acceleration admits a similar decomposition (with a conventional choice of prefactors) 
\bea
\label{airr}
a^i_{\rm rr}&=& -\frac85 \nu \frac{G^2}{c^5} \frac{M^2}{r^3}\left[-A_{\rm rr}\dot r n^i+B_{\rm rr} v^i\right]\,,
\eea
with
\bea
A_{\rm rr}&=&A_{2.5\rm PN}+\eta^2 A_{3.5\rm PN}
\,,\nonumber\\
B_{\rm rr}&=&B_{2.5\rm PN}+\eta^2 B_{3.5\rm PN}\,.
\eea 
The quantities $A_{\rm nPN}$ and $B_{\rm nPN}$ all have a standard PN expansion in powers of $v^2$, $\dot r^2$, $GM/r$, and their products with arbitrary coefficients.
The fundamental scalars in the Hamiltonian formalism are instead $(p^2,p_r^2,GM/r)$.

Following the balance approach the values of these coefficients are partially fixed by assuming that there is a balance between the energy and angular momentum losses of the system in the near zone, and the corresponding radiated energy and angular momentum fluxes at infinity in the form of gravitational waves, which we will denote below by $\Phi_E$ and $\Phi_J$, respectively.
The latter are known gauge-invariant quantities (see, e.g., Eqs. (2.10a)--(2.10b) and Eqs. (2.17a)--(2.17b) in Ref. \cite{Iyer:1995rn} for the corresponding LO and NLO expressions).
The balance equations must also take into account the presence of Schott terms, which represent additional contributions to the energy and angular momentum of the system due to its interaction with the radiation field.
The final set of equations then reads
\bea
\label{balance_laws}
\frac{dE_{\rm sys}}{dt}+\frac{dE_{\rm Schott}}{dt}&=& -\Phi_E\,,\nonumber\\
\frac{dJ_{\rm sys}}{dt}+\frac{dJ_{\rm Schott}}{dt}&=& -\Phi_J\,,
\eea
where $E_{\rm sys}$ and $J_{\rm sys}$ denote the energy and angular momentum of the system, respectively, and $E_{\rm Schott}$ and $J_{\rm Schott}$ the associated Schott terms.
The addition of Schott terms thus introduces a further set of undetermined coefficients, since they admit a similar decomposition as for $A_{\rm nPN}$ and $B_{\rm nPN}$ in terms of the associated fundamental scalars.
As a result the balance equations are not enough to uniquely fix all the unknown coefficients, so that a residual gauge freedom still remains.

\section{Computing the radiation-reaction force}

Let us briefly review the IW balance approach as well as the  BD  Hamiltonian formalism for the computation of the  rr  force.

\subsection{IW approach}

Following the notation of IW \cite{Iyer:1995rn} the rr acceleration is given by Eq. \eqref{airr} with coefficients (denoted by Latin letters) parametrized as 
\bea
A_{2.5\rm PN}&=& a_1 v^2+a_2 \frac{ G M}{r} +a_3 \dot r^2\,,\nonumber\\ 
B_{2.5\rm PN}&=& b_1 v^2+b_2 \frac{ G M}{r} +b_3 \dot r^2\,.
\eea
and
\bea
A_{3.5\rm PN}&=& c_1 v^4 +c_2 v^2 \frac{GM}{r}+c_3 v^2 \dot r^2 +c_4 \dot r^2 \frac{GM}{r}\nonumber\\
&+&c_5 \dot r^4 +c_6 \left(\frac{GM}{r}\right)^2 \nonumber\\
B_{3.5\rm PN}&=&  d_1 v^4 +d_2 v^2 \frac{GM}{r}+d_3 v^2 \dot r^2 +d_4 \dot r^2 \frac{GM}{r}\nonumber\\
&+& d_5 \dot r^4 +d_6 \left(\frac{GM}{r}\right)^2\,.
\eea

The conserved energy and angular momentum of the system through the 2PN order, here denoted by 
\beq
E_{\rm sys}=E_{\rm cons}(x^i,v^i)\,,\qquad
J_{\rm sys}=J_{\rm cons}(x^i,v^i)\,,
\eeq
respectively, are listed in Eqs. (2.3a)--(2.3b) and Eqs. (2.4a)--(2.4f) of Ref. \cite{Iyer:1995rn}.
In the presence of a rr force the energy and angular momentum are no longer conserved.
Therefore, Iyer and Will introduce the quantities
\beq
\tilde E^*=\tilde E_{\rm cons}+\tilde E_{5/2}+\tilde E_{7/2}\,,\qquad 
\tilde J^*=\tilde J_{\rm cons}+\tilde J_{5/2}+\tilde J_{7/2}\,, 
\eeq
where a tilde means per unit of reduced mass $\mu$ of the system, which can be identified with the (gauge-invariant) sum of the energy and angular momentum stored in the system and the Schott energy and angular momentum associated with the gravitational field, namely
\beq
E^*=E_{\rm sys}+E_{\rm Schott}\,,\qquad 
J^*=J_{\rm sys} +J_{\rm Schott}\,, 
\eeq
with
\beq
\label{EJschIW}
E_{\rm Schott}=\mu(\tilde E_{5/2}+\tilde E_{7/2})\,,\qquad 
J_{\rm Schott}=\mu(\tilde J_{5/2}+\tilde J_{7/2})\,.
\eeq
The Schott terms are then expressed in terms of gauge parameters (denoted by Greek letters) as follows
\bea
\tilde E_{5/2}&=&-\frac85 \frac{\nu}{c^5} \left(\frac{GM}{r}\right)^2\dot r\left(\gamma v^2+\delta \frac{ G M}{r} +\beta \dot r^2\right)
\,,\nonumber\\
\tilde E_{7/2}&=&-\frac85 \frac{\nu}{c^7} \left(\frac{GM}{r}\right)^2\dot r\left(\delta_1 v^4 +\delta_2 v^2 \dot r^2 +\delta_3 v^2 \frac{GM}{r}\right.\nonumber\\
&+&\left. 
\delta_4 \dot r^4+\delta_5 \dot r^2 \frac{GM}{r} +\delta_6 \left(\frac{GM}{r}\right)^2\right)
\,,
\eea
and
\bea
\tilde J_{5/2}&=&\frac85 \frac{\nu}{c^5} \frac{GM}{r}\,r^2\dot\phi\,\dot r
   \left(\kappa v^2+\alpha \frac{ G M}{r} +\epsilon \dot r^2\right)
\,,\nonumber\\
\tilde J_{7/2}&=&-\frac85 \frac{\nu}{c^7} \frac{GM}{r}\,r^2\dot\phi\,\dot r
   \left(\epsilon_1 v^4 +\epsilon_2 v^2 \dot r^2 +\epsilon_3 v^2 \frac{GM}{r}\right.\nonumber\\
&+& \left.\epsilon_4 \dot r^4+\epsilon_5 \dot r^2 \frac{GM}{r} +\epsilon_6 \left(\frac{GM}{r}\right)^2\right)\,,
\eea
with $r^2\dot\phi=[{\mathbf x}\times{\mathbf v}]\cdot e_z$.

The balance equations \eqref{balance_laws} thus write as
\beq
\frac{dE^*}{dt}=-\Phi_E\,,\qquad
\frac{dJ^*}{dt} =-\Phi_{J}\,,
\eeq
where  
\bea
\frac{dE^*}{dt}&=&\frac{\partial E_{\rm cons}}{\partial v^i}\,a_{\rm rr}^i +\frac{dE_{\rm Schott}}{dt}\,, \nonumber\\
\frac{dJ^*}{dt}&=&\frac{\partial J_{\rm cons}}{\partial v^i}\,a_{\rm rr}^i +\frac{dJ_{\rm Schott}}{dt}\,.
\eea
The first term in the rhs of the first equation represents the instantaneous variation of the energy of the system, since 
\bea
\frac{dE_{\rm sys}}{dt}&=& \frac{\partial E_{\rm cons}}{\partial x^i}\dot x^i +\frac{\partial E_{\rm cons}}{\partial v^i}a^i\nonumber\\
&=& \frac{\partial E_{\rm cons}}{\partial x^i}\dot x^i +\frac{\partial E_{\rm cons}}{\partial v^i}(a^i_{\rm cons}+a^i_{\rm rr})\nonumber\\
&=&  \frac{\partial E_{\rm cons}}{\partial v^i} a^i_{\rm rr}\,.
\eea
Similar considerations hold for the angular momentum in the second equation, so that 
\beq
\frac{dJ_{\rm sys}}{dt}=\frac{\partial J_{\rm cons}}{\partial v^i}\,a_{\rm rr}^i\,.
\eeq

The balance equations then provide a set of relations between acceleration coefficients and gauge parameters.
At the 2.5PN level one gets
\bea
A_{2.5\rm PN}&=& 3(1+\beta)v^2+\left(\frac{23}{3}+2\alpha-3\beta \right)\frac{GM}{r}-5\beta \dot r^2
\,,\nonumber\\
B_{2.5\rm PN}&=& (2+\alpha)v^2 +(2-\alpha)\frac{GM}{r}-3(1+\alpha)\dot r^2
\,,
\eea
and
\bea
\label{EJschIW52}
\tilde E_{5/2}&=&-\frac85 \frac{\nu}{c^5} \left(\frac{GM}{r}\right)^2\dot r\left(-(\alpha+2) v^2 +\beta \dot r^2\right)
\,,\nonumber\\
\tilde J_{5/2}&=&\frac85 \frac{\nu}{c^5} \left(\frac{GM}{r}\right)^2 r^2\dot\phi\dot r \alpha 
\,,
\eea
The 3.5PN coefficients $c_i$ and $d_i$ in terms of gauge parameters are listed in Table \ref{tab:table35rr}, and 
\bea
\label{EJschIW72}
\tilde E_{7/2}&=&-\frac85 \frac{\nu}{c^7} \left(\frac{GM}{r}\right)^2\dot r\left[\delta_1 v^4 +\delta_2 v^2 \dot r^2 +\delta_3 v^2 \frac{GM}{r}\right.\nonumber\\
&+&\left. 
\delta_4 \dot r^4+\delta_5 \dot r^2 \frac{GM}{r} +\left(\frac{16}{21}\nu-\frac{4}{21}\right) \left(\frac{GM}{r}\right)^2\right]
\,,\nonumber\\
\tilde J_{7/2}&=&-\frac85 \frac{\nu}{c^7} \frac{ GM }{r }r^2\dot\phi\dot r\left[\left(-\frac{137}{21}\nu+\delta_1+\frac{307}{84}\right)v^2 \frac{GM}{r}\right.\nonumber\\
&+&\left. \epsilon_5 \dot r^2 \frac{GM}{r} +\left(\frac{107}{21}\nu+\delta_3-\frac{271}{42}\right) \left(\frac{GM}{r}\right)^2\right]
\,.\nonumber\\
\eea

% table 1

\begin{table}  
\caption{\label{tab:table35rr}  Values of the coefficients $c_i$ and $d_i$ entering the radiation-reaction relative acceleration at the 3.5PN level of accuracy. $\alpha$ and $\beta$ are the two 2.5PN gauge parameters, whereas $\delta_1,\ldots, \delta_5$ and $\epsilon_5$ are the six 3.5PN gauge parameters according to the notation of Ref. \cite{Iyer:1995rn}.}
\begin{ruledtabular}
\begin{tabular}{l|l}
$c_1$ & $\frac{117}{28} +  \frac{33}{7}\nu  -  \frac{3}{2}\beta(1 - 3\nu)  + 3\delta_2 - 3\epsilon_5$   \\ 
$c_2$ & $-\frac{99}{14} +  \frac{155}{21}\nu  - 3\alpha(1 - 4\nu) -  \frac{3}{2}\beta (7 + 13\nu)$ \\
& $- 2\delta_1 - 3\delta_2 + 3\delta_5 + 3\epsilon_5$ \\
$c_3$ & $\frac{95}{28} -  \frac{90}{7}\nu  +  \frac{5}{2}\beta (1 - 3\nu) - 5\delta_2 + 5\delta_4 + 5\epsilon_5$   \\ 
$c_4$ & $-\frac{687}{28} +  \frac{92}{7}\nu   - 6\alpha\nu + \frac{\beta}{2} (54 + 17\nu)$ \\
& $ - 2\delta_2 - 5\delta_4 - 6\delta_5$   \\ 
$c_5$ & $-7\delta_4$   \\ 
$c_6$ & $-73 -  \frac{166}{7}\nu  - \alpha (14 + 9\nu) + 3\beta (7 + 4\nu)$ \\
& $ - 2\delta_3 - 3\delta_5$   \\ 
\hline
$d_1$ & $ -3 + 9\nu -  \frac{3}{2}\alpha (1 - 3\nu)  - \delta_1$   \\ 
$d_2$ & $-\frac{139}{84} -  \frac{64}{7}\nu  - \frac{\alpha}{2} (5 + 17\nu) + \delta_1 - \delta_3$   \\
$d_3$ & $\frac{369}{28} -  \frac{156}{7}\nu  + \frac{3}{2} (3\alpha + 2\beta) (1 - 3\nu)  + 3\delta_1 - 3\epsilon_5$   \\ 
$d_4$ & $\frac{295}{42} -  \frac{335}{42}\nu  + \frac{\alpha}{2} (38 - 11\nu)  - 3\beta (1 - 3\nu)$ \\
& $ + 2\delta_1 + 4\delta_3 + 3\epsilon_5$   \\ 
$d_5$ & $\frac{95}{28} -  \frac{90}{7}\nu  - 5\beta (1 - 3\nu) + 5\epsilon_5$   \\ 
$d_6$ & $-\frac{634}{21} +  \frac{22}{7}\nu  + \alpha (7 + 3\nu) + \delta_3$   \\ 
\end{tabular}
\end{ruledtabular}
\end{table}

The final expression for the rr acceleration thus depends on two gauge parameters, $\alpha$ and $\beta$, at the 2.5PN level, and on six more parameters, $\delta_1$, $\delta_2$, $\delta_3$, $\delta_4$, $\delta_5$, $\epsilon_5$, at the 3.5PN level.

When using harmonic coordinates such parameters are fixed as \cite{Damour:1981bh,Pati:2002ux,Nissanke:2004er}
\bea
&&\alpha=-1\,, \quad \beta=0
\,,\nonumber\\
&&\delta_1=\frac{271}{28}+6\nu\,, \quad \delta_2=-\frac{77}{4}-\frac32 \nu\,, \quad \delta_3=\frac{79}{14}-\frac{92}{7}\nu\,, \nonumber\\
&&\delta_4=10\,, \quad \delta_5=\frac{5}{42}+\frac{242}{21}\nu\,, \quad \epsilon_5=-\frac{439}{28}+\frac{18}{7}\nu
\,.\nonumber\\
\eea
The radial and azimuthal components of the rr force ${\mathcal F}_r^{\rm rr}={\mathcal F}^{\rm rr}\cdot{\mathbf n}$ and ${\mathcal F}_\phi^{\rm rr}=({\mathbf x}\times{\mathcal F}^{\rm rr})\cdot e_z$ are then given by 
\begin{widetext}
\bea
\label{Frphiharm}
{\mathcal F}_r^{\rm rr,h}  & =& 
\frac85\nu^2\eta^5\frac{G^2M^3}{r^3}\dot r\Big\{2v^2+\frac83\frac{GM}{r}
+\eta^2\Big[\Big(\tfrac{65}{14}-6\nu\Big)v^4+\Big(\tfrac{35}{3}\nu-\tfrac{362}{21}\Big)v^2\frac{GM}{r}
+\Big(6\nu-\tfrac{27}{2}\Big)v^2\dot r^2\nonumber\\
&-&\Big(\tfrac{59}{3}+\tfrac{35}{3}\nu\Big)\dot r^2\frac{GM}{r}+5\dot r^4-\Big(\tfrac{821}{21}+10\nu\Big)\left(\frac{GM}{r}\right)^2\Big]\Big\}
\,,\nonumber\\
{\mathcal F}_\phi^{\rm rr,h}&=&  -\frac{8}{5}\nu^2 \eta^5\frac{ G^2M^3}{r} \dot{\phi}\left\{v^2+3\frac{ GM }{r}+\eta^2\left[\left(-\frac{ 313 }{28}-\frac{3}{2}\nu\right)v^4+\left(\frac{205}{42}+\frac{37}{2}\nu\right)v^2 \frac{ GM}{r} +\left(\frac{339}{4}+\frac{3}{2}\nu\right)v^2\dot{r}^2\right.\right. \nonumber\\
&+&\left.\left. 
\left(-\frac{205}{12}-\frac{106}{3}\nu\right)\dot{r}^2\frac{ GM }{r}-75\dot{r}^4
+ \left(-\frac{1325}{42}-13\nu\right)\left(\frac{GM}{r}\right)^2\right]  \right\}\,.
\eea

\end{widetext}

\subsection{BD approach}

The Bini-Damour balance approach starts from the equations of motion of the binary system in the presence of a rr force written in the form \eqref{eq_of_mot0}.
The relative motion within the orbital plane in the cm frame can be conveniently described by using a system of polar coordinates $x^i=(r,\phi)$ with conjugate momenta $p_i=(p_r,p_\phi)$.  
After removal of the rest energy the Hamiltonian gives the energy of the system, while the angular momentum of the system is the $\phi$ component of the momentum
\bea
E_{\rm sys}&=& H(x,p)-M c^2\,,\nonumber\\
J_{\rm sys}&=& p_\phi\,.
\eea 
Assuming that the Hamiltonian does not depend on the azimuthal variable $\phi$, the presence of the rr force is responsible for the following energy and angular momentum losses
\bea
\label{sys_dynamics}
\frac{dE_{\rm sys}}{dt}&=&\dot x^i \frac{\partial H(x,p)}{\partial x^i}+\dot p_i \frac{\partial H(x,p)}{\partial p_i}
=\dot x^i {\mathcal F}_i\,, \nonumber\\
\frac{dJ_{\rm sys}}{dt}&=& -\frac{\partial H(x,p)}{\partial \phi}+ {\mathcal F}_\phi=  {\mathcal F}_\phi\,,
\eea
where we have used Eqs. \eqref{eq_of_mot0}.

The balance equations \eqref{balance_laws} thus write as
\bea
\label{balance_laws2}
\dot r {\mathcal F}_r+\dot\phi{\mathcal F}_\phi+\frac{dE_{\rm Schott}}{dt}&=& -\Phi_E\,,\nonumber\\
{\mathcal F}_\phi+\frac{dJ_{\rm Schott}}{dt}&=& -\Phi_J\,.
\eea
BD constructive algorithm for determining the two components of the radiation-reaction force starts from the results for the energy and angular momentum fluxes, $\Phi_E$ and $\Phi_J$, available in the literature in terms of harmonic relative coordinates and velocities, transforming then these expressions in terms of EOB coordinates and momenta.
The Schott terms are assumed to have a general expressions in terms of EOB coordinates and momenta with undetermined coefficients.
Imposing the vanishing of $J_{\rm Schott}$ immediately gives the azimuthal component ${\mathcal F}_\phi=-\Phi_J$ of the radiation-reaction force. The radial component instead still depends on the remaining gauge parameters associated with $E_{\rm Schott}$. However, making a \lq\lq minimal choice" of these parameters is always possible completely determining ${\mathcal F}_r$ too. We will turn to this issue in Section V.

It is worth to stress that the energy and angular momentum balance equations \eqref{balance_laws2} do not uniquely determine the
instantaneous components of the radiation-reaction force, because of the freedom to include/subtract a total time derivative in the Schott terms and subtracting/adding it to the force. 
Consider then a modification of the force,
\beq
{\mathcal F}_r\rightarrow {\mathcal F}_r+\delta {\mathcal F}_r\,,\qquad 
{\mathcal F}_\phi\rightarrow {\mathcal F}_\phi+\delta {\mathcal F}_\phi\,,
\eeq
accompanied by redefinitions of the Schott contributions,
\bea
E_{\rm Schott}&\rightarrow& E_{\rm Schott}+\delta E_{\rm Schott}\,,\nonumber\\
J_{\rm Schott} &\rightarrow& J_{\rm Schott}+\delta J_{\rm Schott}\,,
\eea
where $\delta E_{\rm Schott}=\delta E_{\rm Schott}(r,\phi,p_r,p_\phi)$ and $\delta J_{\rm Schott}=\delta J_{\rm Schott}(r,\phi, p_r,p_\phi)$ are local functions on phase space.
The balance equations require
\bea
0&=&\dot r\,\delta {\mathcal F}_r+\dot\phi\,\delta {\mathcal F}_\phi+\frac{d}{dt} \delta E_{\rm Schott}  
\,,\nonumber\\
0&=&\delta {\mathcal F}_\phi+\frac{d}{dt}\delta J_{\rm Schott}\,.
\eea
To the leading order in the radiation reaction, the derivatives of $\delta E_{\rm Schott}$ and $\delta J_{\rm Schott}$ are evaluated along the conservative Hamiltonian dynamics.
Consequently, the general local redefinitions of the radiation-reaction force can be expressed as
\bea
\label{deltaF}
\delta {\mathcal F}_r&=&\frac{\dot\phi\,\{\delta J_{\rm Schott},H\}-\{\delta E_{\rm Schott},H\}}{\dot r}\,,\nonumber\\
\delta {\mathcal F}_\phi&=&-\{\delta J_{\rm Schott},H\}\,,
\eea
where
\bea
\label{PB}
\{f,g \}=\sum_i \left(\frac{\partial f}{\partial x^i}\frac{\partial g}{\partial p_i}-\frac{\partial f}{\partial p_i}\frac{\partial g}{\partial x^i}\right)\,,
\eea
denote the Poisson's brackets (hereafter we will assume that the sum over repeated indices is implicit).

\subsection{This work}

The method used in the present work to determine the rr force in a general coordinate system combines IW results valid in harmonic coordinates with the general BD Hamiltonian approach.
We start from the IW balance equations written in harmonic coordinate and their solutions, leading to a reduced set of gauge parameters.
We then translate these results into a Hamiltonian framework by identifying the quantities
\bea
\label{Frphidefs}
\frac{\partial E_{\rm cons}}{\partial v^i}\,a_{\rm rr}^i\bigg\vert_{h\to X}&=&\dot r {\mathcal F}^{\rm rr}_r+\dot\phi{\mathcal F}^{\rm rr}_\phi\bigg\vert_{X}
\,,\nonumber\\
\frac{\partial J_{\rm cons}}{\partial v^i}\,a_{\rm rr}^i\bigg\vert_{h\to X}&=&{\mathcal F}^{\rm rr}_\phi\bigg\vert_{X}
\,,
\eea
written in a different coordinate system $X$.
To do this we only need the coordinate transformation between harmonic coordinates $(r_h,\dot r_h, \phi_h, \dot\phi_h)$ and coordinates and momenta $(r,p_r,\phi,p_\phi)$ in the new coordinate system without solving again the balance equations as in the BD approach. 
Obviously Eqs. \eqref{Frphidefs} do not provide a unique definition for the force components due to the presence of Schott terms in the balance, so that one is always allowed for a convenient reshuffling of contributions upon transforming $E_{\rm Schott}$ and $J_{\rm Schott}$ and taking their time derivative.
In fact, the transformed radial and azimuthal components of the rr force can be directly related to the harmonic ones at the LO only, since
\bea
\frac{\partial E_{\rm cons}}{\partial v^i}\,a_{\rm rr}^i&=&{\mathbf v}\cdot{\mathcal F}_{\rm rr}[1+O(\eta^2)]
\,,\nonumber\\
\frac{\partial J_{\rm cons}}{\partial v^i}\,a_{\rm rr}^i&=&({\mathbf x}\times{\mathcal F}_{\rm rr})^z[1+O(\eta^2)]
\,.
\eea
The energy and angular momentum fluxes, instead, are gauge-invariant (or better \lq\lq equivariant") functions, so that to obtain their expressions in the new coordinate system it is enough to apply the mapping without any ambiguity. 

Having at hand the expressions for the components of the rr force in the new coordinate system one can then solve the associated equations of motion to obtain the solution for the rr correction to the orbit.
However, once this solution is known in harmonic coordinates, one can simply invert the above coordinate transformation to get the solution for the orbit in the new coordinate system as we will show in Section VI.

\section{Passing from one coordinate set to a different one with radiation reaction solutions}

To find the 3.5PN accurate mapping between MH coordinates and either ADM or EOB coordinates we will proceed as in Ref. \cite{Gamboa:2024imd}, where the transformation between harmonic and EOB coordinates was obtained up to the 3PN level, generalizing the method outlined in Ref. \cite{Damour:2000ni} by including radiation reaction.

The starting point is the Hamiltonian description recalled above, which can be equivalently summarized by the following evolution equations 
\bea
\label{new_eq_parentesi_poiss}
\dot r&=& \{r,H\}=\frac{\partial H}{\partial p_r}\,,\nonumber\\
\dot \phi&=& \{\phi,H\}=\frac{\partial H}{\partial p_\phi}\,,\nonumber\\
\dot p_r &=& \{p_r,H\}+{\mathcal F}_r=-\frac{\partial H}{\partial r}+{\mathcal F}_r\,,\nonumber\\
\dot p_\phi &=& \{p_\phi,H\}+{\mathcal F}_\phi=-\frac{\partial H}{\partial \phi}+{\mathcal F}_\phi\,,
\eea
where the components of the external force ${\mathcal F}_i={\mathcal F}_i(x,p)$ are assumed to depend on the phase-space coordinates, and the Poisson brackets \eqref{PB} are used to define the canonical condition for the conjugate variables
\beq
\{r,p_r\}=1\,,\qquad \{\phi,p_\phi\}=1\,.
\eeq
One can also express the last two equations of Eq. \eqref{new_eq_parentesi_poiss} directly in terms of second temporal derivatives of the coordinates instead of first derivatives of the conjugate momenta (within any given phase-space coordinate system)
\bea
\label{new_eq_parentesi_poiss2}
\ddot r &=& \{\{r,H\},H\}+\frac{\partial^2 H}{\partial p_r^2}{\mathcal F}_r+ \frac{\partial^2 H}{\partial p_r\partial p_\phi}{\mathcal F}_\phi\,,  \nonumber\\ 
\ddot \phi &=& \{\{\phi,H\},H\}+\frac{\partial^2 H}{\partial p_r\partial p_\phi}{\mathcal F}_r+ \frac{\partial^2 H}{\partial p_\phi^2}{\mathcal F}_\phi\,, 
\eea
namely
\beq
\ddot x^i =\{\{x^i,H\},H\}+Q^{ij}{\mathcal F}_j\,,
\eeq
where
\beq
Q^{ij}=\begin{pmatrix}
\frac{\partial^2 H}{\partial p_r^2} & \frac{\partial^2 H}{\partial p_r\partial p_\phi} \\
 \frac{\partial^2 H}{\partial p_r\partial p_\phi} & \frac{\partial^2 H}{\partial p_\phi^2}\\
\end{pmatrix}
\eeq
denotes the Hessian matrix of the Hamiltonian with respect to the momenta.
Consequently, as a general result 
\bea
\label{eq:ddotx}
\ddot x^i =\ddot x^i_{\rm cons}+\ddot x^i_{\rm rr}=\{\{x^i,H\},H\}+Q^{ij}{\mathcal F}_j\,,
\eea
where
\bea
\label{accvsforce}
\ddot x^i_{\rm cons}&=& \{\{x^i,H\},H\}\,,\nonumber\\
\ddot x^i_{\rm rr}&=&Q^{ij}{\mathcal F}_j\,,
\eea
where we recall that the $x^i$ are not Cartesian coordinates in general. In other words, $\ddot x^i_{\rm rr}$ denotes an acceleration term within the Hamiltonian formalism, and it is not connected to the corresponding force term by a simple multiplication with the reduced mass $\mu$: the Hessian matrix with respect to the momenta matters here, and plays the role of an inverse effective mass. 

Let us consider the passage from harmonic coordinates $(x^i_h,p^{h}_k)\equiv (r_h,\phi_h; \dot r_h,\dot \phi_h)$ to a new Hamiltonian coordinate system with phase-space variables $(x^i,p_k)\equiv(r,\phi; p_r,p_\phi)$.
The associated coordinate transformation  $x^i_h=f^{i}(x^j,p_k)$ reads explicitly 
\beq
\label{r_h_phi_h}
r_h=f^r\underbrace{(r,\phi; p_r,p_\phi)}_{\rm new\,\, coords}\,,\qquad \phi_h=f^\phi\underbrace{(r,\phi; p_r,p_\phi)}_{\rm new\,\, coords}\,,
\eeq
has a PN-expanded form (up to $O(\eta^7)$) which includes all possible combinations of $(p_r,p_\phi=L,GM/r)$ and their products allowed by dimensional analysis at each order with unknown coefficients. 
For instance, at the leading order one has
\beq
r_h = r[1+\eta^2{\mathcal P}_2+O(\eta^4)]\,,
\eeq
with
\beq
{\mathcal P}_2=k_1 p_r^2 +k_2\frac{L^2}{r^2}+k_3\frac{GM}{r}+k_4 p_r\frac{L}{r}\,.
\eeq
The coefficients are determined by solving a system of linear equations at each PN order obtained by matching the acceleration computed employing Poisson brackets, Eq. \eqref{new_eq_parentesi_poiss2}, with the known expression of the relative acceleration in MH coordinates, whose rr part contains the whole set of unspecified gauge parameters, i.e.,
\beq
\dot x_h^i =\{f^i,H\}+\frac{\partial f^i}{\partial p_j}{\mathcal F}_j\,,
\eeq
namely
\bea
\label{eqs_dots}
\dot r_h &=& \{f^r,H\}+ \frac{\partial f^r}{\partial p_r}{\mathcal F}_r+\frac{\partial f^r}{\partial p_\phi}{\mathcal F}_\phi\,,  \nonumber\\ 
\dot \phi_h &=& \{f^\phi,H\}+  \frac{\partial f^\phi}{\partial p_r}{\mathcal F}_r+\frac{\partial f^\phi}{\partial p_\phi}{\mathcal F}_\phi\,.
\eea
The latter equations once differentiated with  respect to time imply large expressions which are conveniently performed within the explicit cases considered. For example,
\begin{align}
\label{eq:d2f_full}  
\frac{d^2 f^r}{dt^2} ={}& \bigl\{\{f^r, H\}, H\bigr\}
  + \mathcal{F}_i \left[ 2\left\{ \frac{\partial f^r}{\partial p_i}, H \right\}
  + \bigl\{ f^r, \dot{x}^i \bigr\} \right] \nonumber \\
&+ \{\mathcal{F}_i, H\}\, \frac{\partial f^r}{\partial p_i}
  + \mathcal{F}_j\, \frac{\partial}{\partial p_j}
  \left( \mathcal{F}_i\, \frac{\partial f^r}{\partial p_i} \right),
\end{align}
where $x^i = (r, \phi)$, $p_i = (p_r, p_\phi)$,
$\dot{x}^i = \partial H / \partial p_i$.
Working at the linear order in the radiation-reaction force (as well as in the present case at the 3.5PN approximation level) the above relation simplifies as
\bea
\label{eq:d2f_linear}
\frac{d^2 f^r}{dt^2} &=& \bigl\{\{f^r, H\}, H\bigr\}
  + \mathcal{F}_i \left[ 2\left\{ \frac{\partial f^r}{\partial p_i}, H \right\}
  + \left\{ f^r, \frac{\partial H}{\partial p_i} \right\} \right]\nonumber\\
&+& \{\mathcal{F}_i, H\}\, \frac{\partial f^r}{\partial p_i}+O({\mathcal F}^2)\,.
\eea
The left-hand side    reproduces well-known relations
\bea
\label{MHacc}
\ddot r_h &=&r_h\dot \phi_h^2+a^r(r_h,\phi_h; \dot r_h,\dot \phi_h)\,,  \nonumber\\ 
\ddot \phi_h &=&-\frac{2\dot r_h\dot \phi_h}{r_h}+a^\phi(r_h,\phi_h; \dot r_h,\dot \phi_h)\,,
\eea
where $a^r={\mathbf a}\cdot {\mathbf n}$ and $ra^\phi={\mathbf a}\cdot e_\phi$ (with $e_\phi$ the unit vector along the azimuthal direction) denote the radial and azimuthal components of the (conservative plus rr) acceleration in MH coordinates.  
We emphasize that all quantities at the right-hand side of the above equations,  Eq. \eqref{MHacc}, depend on the \lq\lq new" phase space variables $(r,\phi, p_r,p_\phi)$, so that one has to substitute the mapping \eqref{r_h_phi_h} as well as its time derivative \eqref{eqs_dots}.
Hereafter, we will assume that the external force coincides with the rr force. 

The resulting transformations between MH coordinates and either ADM or EOB coordinates are listed in Appendix A.

\section{Radiation-reaction force within various coordinate systems}

The components of the rr force in the new Hamiltonian coordinate system $X$ directly follow from Eq. \eqref{Frphidefs}, where the lhs of each equation (written in harmonic coordinates) is evaluated by applying the mapping $h\to X$ derived in the previous section.
The Schott energy and angular momentum are instead simply obtained by transforming the coordinates in their (harmonic-coordinate) definition, Eqs. \eqref{EJschIW} and \eqref{EJschIW52}--\eqref{EJschIW72}.

We stress that the relations between the coefficients $c_i$, $d_i$ of the radiation-reaction acceleration and the gauge
parameters listed in Table \ref{tab:table35rr} follow from the balance equations written with the harmonic-coordinate conserved energy and
angular momentum, energy and angular momentum fluxes, and 1PN conservative dynamics. They therefore hold in any coordinate
system that agrees with harmonic coordinates through 1PN order, as is the case for ADM coordinates, but not in EOB
coordinates, which differ from harmonic ones already at 1PN. As a consequence, an acceleration of the IW form with the
coefficients of Table \ref{tab:table35rr}, written in EOB variables and converted into a force through the Hessian of $H_{\rm EOB}$ with
respect to the momenta, does not satisfy the balance equations at 3.5PN for any choice of the gauge parameters, nor can it
be obtained from the harmonic-coordinate dynamics through a change of phase-space variables. The EOB force must be built
either through Eq. \eqref{Frphidefs}, as done here, or by re-deriving the analogue of Table~II in EOB coordinates; the two
procedures give the same eight-parameter family of forces, up to a relabeling of the 3.5PN gauge parameters.

We list below the resulting expressions for the radiation-reaction force components and the Schott terms in both ADM and EOB coordinate systems.
Hereafter, we will set $G=1$ and $M=1$ for simplicity. 
We also avoid specifying the coordinate system under consideration with an extra label to lighten the notation.

\subsection{ADM}

In the ADM case the gauge parameters are fully determined.
The Schott contributions to the energy and angular momentum are simply obtained by transforming the corresponding harmonic expressions with the ADM values of the gauge parameters, leading to 

\begin{widetext}

\bea
E_{\rm Schott}&=&\frac{8}{15}\eta^5\nu^2\frac{p_r}{r^2}\left\{
\frac{11L^2}{r^2}+2 p_r^2
+\eta^2 \left[\frac{L^4}{r^4}\left(\frac{8 \nu }{7}-\frac{503}{28}\right)+ \frac{L^2p_r^2}{r^2}\left(-\frac{41 \nu }{7}-\frac{131}{14}\right)
   +\frac{L^2}{r^3}\left(-\frac{151 \nu }{7}-\frac{4521}{28}\right)\right.\right.\nonumber\\
&&\left.\left.
+\left(-\frac{4 \nu }{7}-\frac{26}{7}\right)
   p_r^4+\frac{ p_r^2}{r}\left(-\frac{39 \nu }{7}-\frac{331}{28}\right)+\frac{1}{r^2}\left(\frac{4}{7}-\frac{16 \nu }{7}\right)\right]
\right\}
\,,\nonumber\\
J_{\rm Schott}&=&\frac{8}{15}\eta^5\nu^2\frac{p_rL}{r^2}\left\{
5+\eta ^2 \left[\frac{L^2}{r^2} \left(-\frac{193 \nu }{14}-\frac{122}{7}\right)+\left(-\frac{62 \nu }{7}-\frac{13}{2}\right)
   p_r^2+\frac{1}{r}\left(-\frac{55 \nu }{7}-\frac{2047}{28}\right)\right]
\right\}
\,,
\eea
whereas the radial and azimuthal components of the radiation reaction force are given by Eqs. \eqref{Frphidefs}
\bea
\label{Frradm}
{\mathcal F}^{\rm rr}_r&=&\frac{8}{15}\eta^5\nu^2\frac{p_r}{r^3}\left\{
\frac{25 L^2}{r^2}+4 p_r^2+\frac{5}{r}
+\eta^2 \left[
\frac{L^4}{r^4} \left(\frac{1109 \nu }{14}-\frac{178}{7}\right)+ \frac{L^2p_r^2}{r^2}\left(\frac{125}{14}-\frac{211 \nu }{7}\right)
   +\frac{L^2}{r^3}\left(-\frac{1507 \nu }{14}-\frac{11359}{28}\right)\right.\right.\nonumber\\
&&\left.\left.
+\left(-\frac{8 \nu }{7}-\frac{52}{7}\right)
   p_r^4+\frac{p_r^2}{r}\left(-\frac{373 \nu }{14}-\frac{1035}{28}\right)+\frac{1}{r^2}\left(-\frac{93 \nu }{7}-\frac{725}{28}\right)
\right]
\right\}
\,,\nonumber\\
{\mathcal F}^{\rm rr}_\phi&=&\frac{8}{15}\eta^5\nu^2\frac{L}{r^3}\left\{
-\frac{11L^2}{r^2}+13 p_r^2-\frac{1}{r}
+\eta^2 \left[
\frac{L^4}{r^4} \left(\frac{503}{28}-\frac{8 \nu }{7}\right)+\frac{L^2p_r^2}{r^2}\left(-\frac{767 \nu }{14}-\frac{380}{7}\right)
   +\frac{L^2}{r^3}\left(\frac{169 \nu }{7}+\frac{4017}{28}\right)\right.\right.\nonumber\\
&&\left.\left.
+\left(\frac{67 \nu }{7}-\frac{389}{14}\right)
   p_r^4+\frac{p_r^2}{r}\left(-\frac{1255 \nu }{14}-\frac{8501}{28}\right) +\frac{1}{r^2}\left(-2 \nu -\frac{193}{28}\right)
\right]
\right\}
\,,
\eea
\end{widetext}
respectively.

In the ADM case the components of the rr force can be also directly computed by inverting Eq. \eqref{accvsforce} relating the rr acceleration to the rr force. 
Noticeably, only with $H$ at the Newtonian level (or equivalently with $\ddot x^i_{\rm rr}$ at the leading 2.5PN level)
\beq
H_{\rm N}= \frac{p^2}{2\mu }-\frac{GM\mu }{r}\,,
\eeq
one finds
\beq
Q^{ij}_{\rm N}=\begin{pmatrix}
1 & 0 \\
0 & \frac{1}{r^2}\\
\end{pmatrix}\,,
\qquad 
[\left(Q^{\rm N}\right)^{-1}]_{ij}=\begin{pmatrix}
1 & 0 \\
0 &  r^2 \\
\end{pmatrix}\,,
\eeq
and the well-known relation is re-established
\beq
\ddot x^i_{\rm rr}=Q^{ij}{\mathcal F}_j=\frac{1}{\mu}{\mathcal F}^i=a_{\rm rr\, 2.5PN}^i\,,
\eeq
where one should note the  index $i$ in a contravariant position.
Beyond the Newtonian level for $H$ (namely already at the  3.5PN level which is the one considered here)  the above relation $\ddot x^i_{\rm rr}=a_{\rm rr\, 2.5PN}^i$ is no longer  true, in the sense that it acquires terms involving momenta. Let us illustrate this feature more in detail. 

Consider the ADM Hamiltonian, truncated at 1PN, rescaled by the reduced mass (and in terms of rescaled momenta too $p_i\to p_i/\mu$ and rescaled radial variable $GM/r\to 1/r$)
\bea
\hat H_{\rm ADM}(x,p)&=& \hat H_{\rm N}+\hat H_{\rm 1PN}\\
&=& \frac{p^2}{2}-\frac{1}{r}\nonumber\\
&+&\eta^2 \left[  \frac{3\nu-1}{8}\,p^4 - \frac{(3+\nu)\,p^2 + \nu\,p_r^2}{2r }+ \frac{1}{2r^2} \right]
\,,\nonumber
\eea
where
\beq
p^2 =p_r^2+ \frac{p_\phi^2}{r^2}\,.
\eeq
The Hessian matrix with respect the momenta is given by 
\bea
\frac{\partial^2 \hat H_{\rm ADM}}{\partial p_r^2}&=& 1+\eta^2 \left[\frac{3\nu-1}{2}\left(\frac{p_\phi^2}{r^2}+3p_r^2 \right)-\frac{3+2\nu}{r}\right]\,,\nonumber\\
\frac{\partial^2 \hat H_{\rm ADM}}{\partial p_r\partial p_\phi }&=&\eta^2 (3\nu-1)\frac{p_r p_\phi}{r^2}\,, \nonumber\\
\frac{\partial^2 \hat H_{\rm ADM}}{\partial p_\phi^2}&=& \frac{1}{r^2}+\eta^2\left[\frac{3\nu-1}{2r^2}\left(p_r^2+3\frac{p_\phi^2}{r^2} \right)-\frac{3+\nu}{r^3}\right]\,,\nonumber\\
\eea
that is, in matrix form (and omitting the label ADM hereafter in this section to ease notation)
\bea
Q_{ij}=Q^{\rm N}_{ij}+\eta^2 Q^{\rm 1PN}_{ij}\,,
\eea
with, for example,
\beq
Q^{\rm 1PN}_{11}=\frac{3\nu-1}{2}\left(\frac{p_\phi^2}{r^2}+3p_r^2 \right)-\frac{3+2\nu}{r}\,,
\eeq
etc.
This implies
\bea
\label{da_invertire}
\begin{pmatrix}
\ddot r_{\rm rr} \\
\ddot \phi_{\rm rr}\\
\end{pmatrix}
&=& 
Q^{\rm N}_{ij}
\begin{pmatrix}
F_r^{\rm rr,2.5PN}+\eta^2 F_r^{\rm rr,3.5PN} \\
F_\phi^{\rm rr,2.5PN}+\eta^2 F_\phi^{\rm rr,3.5PN} \\
\end{pmatrix}\nonumber\\
&+& \eta^2 Q^{\rm 1PN}_{ij}
\begin{pmatrix}
F_r^{\rm rr,2.5PN}  \\
F_\phi^{\rm rr,2.5PN} \\
\end{pmatrix}\,.
\eea
Inverting Eq. \eqref{da_invertire} in PN sense (restoring the $\mu$ factors and recalling that $\ddot \phi=a_\phi/r$) leads to
\bea
\label{Frr_vs_arr}
F^{\rm rr,2.5PN}_r &=& \mu \ddot r_{\rm rr}^{\rm 2.5PN} =\mu a_{{\rm rr},r}^{\rm 2.5PN}\,,\nonumber\\
F^{\rm rr,2.5PN}_\phi &=& \mu r^2 \ddot \phi_{\rm rr}^{\rm 2.5PN} =\mu r a_{{\rm rr},\phi }^{\rm 2.5PN} \,,\nonumber\\
F^{\rm rr,3.5PN}_r &=& \mu \ddot r_{\rm rr}^{\rm 3.5PN}\nonumber\\
&-& \eta^2 \mu \left[Q^{\rm 1PN}_{11}a_{{\rm rr},r}^{\rm 2.5PN} + Q^{\rm 1PN}_{12}\frac{a_{{\rm rr},\phi }^{\rm 2.5PN}}{r}\right]\,,\nonumber\\
F^{\rm rr,3.5PN}_\phi &=& \mu r^2 \ddot \phi_{\rm rr}^{\rm 3.5PN}\nonumber\\
&-& \mu r^2 \eta^2 \left[Q^{\rm 1PN}_{12}a_{{\rm rr},r}^{\rm 2.5PN} + Q^{\rm 1PN}_{22} r a_{{\rm rr},\phi }^{\rm 2.5PN} \right]\,,\nonumber\\
\eea
which coincides with Eq. \eqref{Frradm} above.

\subsection{EOB}

As already mentioned, in the EOB case there is no way to uniquely fix the gauge parameters, since the EOB formalism only exists in the cm system. The multi-parameter arbitrariness of the IW construction of a radiation-reaction force by the balance method thus fully remains both in the Schott energy and angular momentum and in the components of the radiation reaction force.

Differently from the ADM case, computing the rr force through Eqs. \eqref{Frphidefs} would lead to a radial component exhibiting unwanted terms behaving like $1/p_r$. These terms can be removed by suitably modifying the definition of the Schott contribution to the energy in the EOB coordinates, whereas the angular momentum is taken to be simply the transformed harmonic expression. 
We finally get
\begin{widetext}
\bea
E_{\rm Schott}&=&\frac{8}{5}\eta^5\nu^2\frac{p_r}{r^2}\left\{
p_r^2 (\alpha -\beta+2)+\frac{L^2}{r^2}(\alpha +2) 
+\eta ^2 \left[
\frac{L^2 p_r^2}{r^2} \left(3 \alpha  \nu -3 \alpha -5 \beta \nu +\frac{3 \beta}{2}-2 \delta_1-\delta_2+6   \nu -6\right)\right.\right.\nonumber\\
&&
+ \frac{L^2}{r^3}(-\nu\alpha -5 \alpha -\delta_3-6 \nu -10)
+\frac{p_r^2}{r} (-5 \alpha  \nu -7 \alpha +5
   \beta \nu +7 \beta-\delta_3-\delta_5-10 \nu -14)\nonumber\\
&&\left.\left.
+p_r^4 \left(-\frac{3 \alpha }{2}+\frac{3 \beta}{2}-\delta_1-\delta_2-\delta_4-3\right)
+\frac{L^4}{r^4} \left(2 \alpha  \nu -\frac{3 \alpha }{2}-\delta_1+4 \nu-3\right)
+\frac{1}{r^2}\left(\frac{4}{21}-\frac{16 \nu }{21}\right)
\right]
\right\}   
\,,\nonumber\\
J_{\rm Schott}&=&\frac{8}{5}\eta^5\nu^2\frac{p_rL}{r^2}\left\{
\alpha
+\eta ^2 \left[
p_r^2 \left(-\frac{\alpha  \nu }{2}-\alpha -\delta_1-\epsilon_5+\frac{137 \nu }{21}-\frac{307}{84}\right)
+\frac{L^2}{r^2}
   \left(\frac{5 \alpha  \nu }{2}-\alpha -\delta_1+\frac{137 \nu }{21}-\frac{307}{84}\right)\right.\right.\nonumber\\
&&\left.\left.
+\frac{1}{r}\left(-2 \alpha  \nu -4 \alpha -\delta_3-\frac{107 \nu }{21}+\frac{271}{42}\right)
\right]
\right\}
\,,
\eea
and
\bea
{\mathcal F}^{\rm rr}_r&=&\frac{8}{5}\eta^5\nu^2\frac{p_r}{r^3}\left\{
\frac{L^2}{r^2} (-\alpha +3 \beta+1)+p_r^2 (2 \alpha -2 \beta+4)+\frac{1}{r}\left(3 \alpha -3 \beta+\frac{17}{3}\right)\right.\nonumber\\
&&
+\eta ^2 \left[
\frac{L^4}{r^4} \left(-8 \alpha  \nu +\frac{3 \alpha }{2}+15 \beta \nu -\frac{9 \beta}{2}+\delta_1+3 \delta_2-3 \epsilon_5-\frac{53 \nu }{14}+\frac{117}{28}\right)\right.\nonumber\\
&&
+\frac{L^2p_r^2}{r^2} \left(14 \alpha  \nu -\frac{3 \alpha }{2}-20 \beta \nu
   -\frac{3 \beta}{2}-\delta_1+\delta_2+5 \delta_4+2 \epsilon_5+\frac{173 \nu }{14}+\frac{11}{7}\right)\nonumber\\
&&
+\frac{L^2}{r^3}\left(25 \alpha  \nu
   +\frac{3 \alpha }{2}-30 \beta \nu -\frac{27 \beta}{2}-3 \delta_1-3 \delta_2+\delta_3+3 \delta_5+3 \epsilon_5+\frac{701
   \nu }{42}-\frac{101}{12}\right)\nonumber\\
&&
+\frac{p_r^2}{r} \left(-15 \alpha  \nu -\frac{45 \alpha }{2}+15 \beta \nu +\frac{45
   \beta}{2}-5 \delta_1-5 \delta_2-3 \delta_3-5 \delta_4-3 \delta_5-\frac{1247 \nu }{42}-\frac{1819}{42}\right)\nonumber\\
&&
+\frac{1}{r^2}\left(-15
   \alpha  \nu -9 \alpha +15 \beta \nu +9 \beta-3 \delta_3-3 \delta_5-\frac{683 \nu }{21}-\frac{141}{7}\right)\nonumber\\
&&\left.\left.
+p_r^4 (-3\alpha +3 \beta-2 \delta_1-2 \delta_2-2 \delta_4-6)
\right]
\right\}
\,,\nonumber\\
{\mathcal F}^{\rm rr}_\phi&=&\frac{8}{5}\eta^5\nu^2\frac{L}{r^3}\left\{
-(\alpha+2) \frac{L^2}{r^2}+(2 \alpha +1) p_r^2+\frac{1}{r}(\alpha -2)\right.\nonumber\\
&&
+\eta ^2 \left[
\frac{L^2 p_r^2}{r^2} \left(11 \alpha  \nu -\frac{3 \alpha }{2}-\delta_1+3 \epsilon_5+\frac{137 \nu
   }{14}-\frac{117}{28}\right)
+\frac{L^2}{r^3} \left(3 \alpha  \nu +\frac{9 \alpha }{2}-\delta_1+\delta_3+\frac{127 \nu
   }{7}+\frac{139}{84}\right)\right.\nonumber\\
&&
+p_r^4 \left(-2 \alpha  \nu -3 \alpha -2 \delta_1-2 \epsilon_5+\frac{163 \nu
   }{14}-\frac{74}{7}\right)
+\frac{p_r^2}{r} \left(-6 \alpha  \nu -\frac{39 \alpha }{2}-3 \delta_1-3 \delta_3-3 \epsilon_5+\frac{202 \nu
   }{21}-\frac{227}{84}\right)\nonumber\\
&&\left.\left.
+\frac{L^4}{r^4} \left(-2 \alpha  \nu +\frac{3 \alpha }{2}+\delta_1-4 \nu +3\right)
+\frac{1}{r^2}\left(-\nu\alpha -3 \alpha -\delta_3-\frac{50 \nu }{7}+\frac{466}{21}\right)
\right]
\right\}  
\,.
\eea
The Schott energy correction needed to remove the $1/p_r$ terms in the radial component is
\beq
\delta E_{\rm Schott}\equiv E_{\rm Schott}-E_{\rm Schott}^{\rm IW}\big|_{h\to{\rm EOB}}
=\frac85\,\eta^7\nu^3\,\frac{p_rL^2}{r^4}\left[\frac{\alpha-2}{r}-(\alpha+2)\frac{L^2}{r^2}\right]\,,
\eeq
so that (see Eq. \eqref{deltaF})
\bea
\delta {\mathcal F}^{\rm rr}_r&\equiv&-\frac{1}{\dot r}\frac{d\,\delta E_{\rm Schott}}{dt}
=-\frac{1}{\dot r}\{\delta E_{\rm Schott},H_{\rm EOB}\}\nonumber\\
&=&\frac85\,\eta^7\nu^3\,\frac{L^2}{r^4}\left[p_r\left(\frac{5 (\alpha -2)}{r^2}-\frac{6 (\alpha +2) L^2}{r^3}\right)
+\frac{1}{p_r}\left(\frac{(\alpha +2) L^4}{r^5}-\frac{2 \alpha  L^2}{r^4}+\frac{\alpha -2}{r^3}\right)\right]\,.
\eea

One can fix part of the undetermined gauge coefficients by performing special choices. We will consider explicitly the BD gauge and several different choices made in Refs. \cite{Khalil:2021txt,Ramos-Buades:2021adz,Gamboa:2024imd}.

\subsubsection{Gauge choices in Ref. \cite{Bini:2012ji}}

The BD gauge consists in imposing the condition $J_{\rm Schott}\equiv0$, namely
\beq
\alpha = 0\,, \qquad 
\delta_1 = \frac{137}{21}\nu-\frac{307}{84}\,, \qquad 
\delta_3 = \frac{271}{42}-\frac{107}{21}\nu\,, \qquad 
\epsilon_5 = 0\,,
\eeq
so that ${\mathcal F}_\phi=-\Phi_J$, i.e.,
\bea
{\mathcal F}^{\rm rr}_\phi&=&\frac{8}{5}\eta^5\nu^2\frac{L}{r^3}\left\{
-\frac{2L^2}{r^2}+p_r^2-\frac{2}{r}
+\eta ^2 \left[
\frac{L^4}{r^4} \left(\frac{53 \nu }{21}-\frac{55}{84}\right)
+\frac{L^2 p_r^2}{r^2}\left(\frac{137 \nu }{42}-\frac{11}{21}\right)  
+\frac{L^2}{r^3}\left(\frac{137 \nu }{21}+\frac{247}{21}\right)\right.\right.\nonumber\\
&&\left.\left.
+\left(-\frac{59 \nu }{42}-\frac{137}{42}\right)p_r^4
+\frac{p_r^2}{r}\left(\frac{16 \nu }{3}-\frac{233}{21}\right) 
+\frac{1}{r^2}\left(\frac{661}{42}-\frac{43 \nu }{21}\right)\right]
\right\} 
\,,
\eea
and
\bea
\label{EschJscheq0}
E_{\rm Schott}&=&\frac{8}{5}\eta^5\nu^2\frac{p_r}{r^2}\left\{ 
(2-\beta) p_r^2+\frac{2 L^2}{r^2}
+\eta^2 \left[
\frac{L^2p_r^2}{r^2} \left(-5 \beta \nu +\frac{3 \beta}{2}-\delta_2-\frac{148 \nu}{21}+\frac{55}{42}\right)
+\frac{L^2}{r^3}\left(-\frac{19 \nu }{21}-\frac{691}{42}\right)\right.\right.\nonumber\\
&&
+p_r^4 \left(\frac{3 \beta}{2}-\delta_2-\delta_4-\frac{137 \nu }{21}+\frac{55}{84}\right)
+\frac{p_r^2}{r} \left(5 \beta \nu +7 \beta-\delta_5-\frac{103 \nu}{21}-\frac{859}{42}\right)\nonumber\\
&&\left.\left.
+\frac{L^4}{r^4} \left(\frac{55}{84}-\frac{53 \nu }{21}\right)
+\frac{1}{r^2}\left(\frac{4}{21}-\frac{16 \nu }{21}\right)
\right]
\right\}
\,,
\eea
and
\bea
\label{FrJscheq0}
{\mathcal F}^{\rm rr}_r&=&\frac{8}{5}\eta^5\nu^2\frac{p_r}{r^3}\left\{
\frac{L^2}{r^2}(3\beta+1) +(4-2 \beta) p_r^2+\frac{1}{r}\left(\frac{17}{3}-3 \beta\right)
+\eta^2\left[
\frac{L^4}{r^4} \left(15 \beta \nu -\frac{9 \beta}{2}+3 \delta_2+\frac{115 \nu }{42}+\frac{11}{21}\right)\right.\right.\nonumber\\
&&
+\frac{L^2p_r^2}{r^2} \left(-20 \beta \nu -\frac{3 \beta}{2}+\delta_2+5 \delta_4+\frac{35 \nu }{6}+\frac{439}{84}\right)
+\frac{L^2}{r^3}\left(-30 \beta \nu -\frac{27 \beta}{2}-3 \delta_2+3 \delta_5-\frac{335 \nu}{42}+9\right)\nonumber\\
&&
+p_r^4 \left(3 \beta-2 \delta_2-2 \delta_4-\frac{274 \nu }{21}+\frac{55}{42}\right)
+\frac{p_r^2}{r}\left(15 \beta \nu +\frac{45 \beta}{2}-5 \delta_2-5 \delta_4-3 \delta_5-\frac{1975 \nu}{42}-\frac{1243}{28}\right)\nonumber\\
&&\left.\left.
+\frac{1}{r^2}\left(15 \beta \nu +9 \beta-3 \delta_5-\frac{362 \nu }{21}-\frac{79}{2}\right)
\right]
\right\}  
\,.
\eea
The request $J_{\rm Schott}=0$ can be complemented by additional choices to fix all the remaining gauge parameters. 

Bini and Damour discussed a possible such choice by noting that the combined flux $\Phi_{EJ}=\Phi_E-\dot\phi\,\Phi_J$ can be  decomposed in a part proportional to $p_r$ (and therefore to $\dot r$), and a total derivative.
One is then free to move part of this flux either to the radial component of the rr force or to the Schott contribution to the energy, since they are related by $\dot r{\mathcal F}_r+\frac{dE_{\rm Schott}}{dt}=-\Phi_{EJ}$.
The BD procedure to obtain \lq\lq minimal" expressions for $E_{\rm Schott}$ and ${\mathcal F}_r$ is rather involved.
However, it can be easily shown that there exist a particular choice of the remaining gauge parameters in our general expressions \eqref{EschJscheq0} and \eqref{FrJscheq0} reproducing the BD minimal solution.
In fact, consider for instance the BD minimal gauge expression for the Schott energy, Eqs. (C1)--(C3) of Ref. \cite{Bini:2012ji},
\bea
E_{\rm Schott}^{\rm (min)\,BD}&=&\eta^5\nu^2\frac{p_r}{r^2}\left\{
\frac{16 p^2}{5}-\frac{16 p_r^2}{5}
+\eta ^2 \left[\left(\frac{22}{21}-\frac{424 \nu }{105}\right) p^4+\left(\frac{136 \nu }{21}-\frac{388}{105}\right)
   p^2p_r^2+\left(-\frac{152 \nu }{105}-\frac{2764}{105}\right)\frac{p^2}{r}\right.\right.\nonumber\\
&&\left.\left.
+\left(\frac{278}{105}-\frac{256 \nu }{105}\right)
   p_r^4+\left(\frac{64 \nu }{21}+\frac{452}{21}\right)\frac{p_r^2}{r}+\left(\frac{32}{105}-\frac{128 \nu
   }{105}\right)\frac{1}{r^2}\right]
\right\}\,.
\eea
Direct comparison with Eq. \eqref{EschJscheq0} shows that the two expressions agree if
\beq
\label{BDminchoice}
\beta=2\,,\qquad
\delta_2 = \frac{223}{42}-\frac{337}{21}\nu\,,\qquad
\delta_4 = -\frac{139}{84}+\frac{200}{21}\nu\,,\qquad
\delta_5 = -\frac{145}{42}+\frac{86}{21}\nu\,.
\eeq
The radial components of the rr force also agree (see Eqs. (D1)--(D3) of Ref. \cite{Bini:2012ji}).

\subsubsection{Gauge choices in Refs. \cite{Khalil:2021txt,Ramos-Buades:2021adz,Gamboa:2024imd}}

Refs. \cite{Khalil:2021txt,Ramos-Buades:2021adz,Gamboa:2024imd} have explored different gauge choices for the radiation-reaction force at the leading and next-to-leading orders, to be used in the construction of EOB-based waveform models.

In Refs. \cite{Khalil:2021txt,Ramos-Buades:2021adz} the authors work directly in EOB coordinates. 
They focused on the quasi-circular case, and specify the gauge coefficients in the Schott terms in such a way that the components of the force satisfy the conditions
\bea
{\mathcal F}_\phi&=&-\Phi_J+O(p_r^2)
\,,\nonumber\\
\frac{{\mathcal F}_rp_\phi}{{\mathcal F}_\phi p_r}&=&1+O(p_r^2)\,.
\eea
We recall that for circular orbits the following relations hold
\beq
L= \sqrt{r}+\frac{3 \eta ^2}{2 \sqrt{r}}-\frac{3 \eta ^4 (4 \nu -9)}{8 r^{3/2}}
+\frac{\eta ^6 \left(123 \pi ^2 \nu -3464 \nu +810\right)}{96 r^{5/2}}\,,
\eeq
and
\beq
x=(M\Omega)^{2/3}= 
\frac{1}{r}+\frac{\eta ^2 \nu }{3r^2}+\frac{\eta ^4 \nu  (8 \nu -45)}{36 r^3}
+\frac{\eta ^6 \nu  \left(224 \nu ^2-432 \nu +1107 \pi ^2-28530\right)}{1296 r^4}
\eeq
with $\Omega=\frac{\partial H}{\partial p_\phi}\big\vert_{\rm circ}$.

Substituting the above expressions for the components of the rr force we find that the first equation is identically satisfied, whereas the second one gives
\bea
\frac{{\mathcal F}_rp_\phi}{{\mathcal F}_\phi p_r}-1&=&
-\frac{\alpha }{2}-\frac{8}{3}
+\frac{\eta ^2 \left(-\frac{11 \alpha  \nu }{8}-\frac{239 \alpha }{672}+\frac{\delta_1 }{2}+\frac{\delta_3}{2}+\frac{167 \nu
   }{84}-\frac{481}{144}\right)}{r}
+O(p_r^2)\,,
\eea
which implies
\beq
\label{betal1}
\alpha=-\frac{16}{3}\,,\qquad
\delta_1 +\delta_3= \frac{485}{168}- \frac{261}{14} \nu\,.
\eeq
The 2.5PN parameter $\beta$ is still arbitrary.

In Ref. \cite{Khalil:2021txt,Ramos-Buades:2021adz} the authors choose $\beta=-1$.
Their final expressions for the force components, Eqs. (40) and (43) in Ref. \cite{Khalil:2021txt}, agree with our results if
\bea
&&\alpha=-\frac{16}{3}\,,\qquad \beta=-1\,,\qquad
\delta_1=\frac{47}{28}-\frac{143}{21}\nu\,,\qquad
\delta_2=-\frac{3}{2}+\frac{25}{3}\nu\,,\qquad
\delta_3=\frac{29}{24}-\frac{71}{6}\nu\,,\nonumber\\
&&\delta_4=-\frac{5}{3}\nu\,,\qquad
\delta_5=\frac{1217}{63}\nu-\frac{29}{4}\,,\qquad
\epsilon_5=\frac{11}{28}+\frac{197}{14}\nu\,.
\eea

In Ref. \cite{Gamboa:2024imd} the authors choose instead $\beta=-13/2$ (with the same value of $\alpha=-{16}/{3}$) in order to have no modification of the orbital phase in the circular orbit limit ($p_r=0$) when transforming the azimuthal angle $\phi$ from harmonic to EOB coordinates at the leading order, where
\beq
\phi_h=\phi+\frac{8\nu}{15}\eta^5(2\beta+13)x^{5/2}+O(\eta^7)\,.
\eeq
Upon substituting $\beta=-13/2$ together with Eq. \eqref{betal1} into our general expressions for the components of the rr force direct comparison with the corresponding expressions obtained in Ref. \cite{Gamboa:2024imd}, Eqs. (66a) and (66b), shows that they both agree if one also sets
\bea
\label{betal2}
&&\delta_1 = \frac{47}{28}-\frac{143}{21}\nu\,, \qquad 
\delta_2 = -\frac{39}{4}+\frac{215}{6}\nu\,, \qquad 
\delta_3 = \frac{29}{24}-\frac{71}{6}\nu\,, \nonumber\\
&&\delta_4 = -\frac{175}{6}\nu\,, \qquad 
\delta_5 = -\frac{233}{6}-\frac{91}{6}\nu\,, \qquad 
\epsilon_5 = 16\nu
\,.
\eea
At the next-to-leading order we find
\bea
\phi_h&=&\phi+
\nu \eta^7 \left(\frac25\delta_1+\frac1{15}\delta_2+\frac35\delta_3-\frac{11}{75}\delta_4+\frac3{10}\delta_5-\frac15\epsilon_5
+\frac{1439}{420}\nu-\frac{1439}{140}\right)x^{7/2}+O(\eta^9)\,, 
\eea
with $\alpha=-{16}/{3}$ and $\beta=-13/2$, so that the further constraint 
\beq
\delta_2=  -\frac{1439}{28}\nu+\frac{4317}{28}+\frac{11}{5}\delta_4 +3\epsilon_5-9\delta_3-\frac{9}{2}\delta_5-6\delta_1
\,,
\eeq
on the 3.5PN gauge parameters extends to the NLO the agreement between the quasi-circular orbital phase in harmonic and EOB coordinates.
However, the above condition is not compatible with the choice of parameters as in Eq. \eqref{betal2}. 

\end{widetext}

\section{Radiation-reaction corrected orbit}

In a previous work \cite{Bini:2025rng} we have computed the radiation-reaction corrected orbit in harmonic coordinates at the 3.5PN order in the case of hyperboliclike motion by using a quasi-Keplerian parametrization for the orbit and the method of variation of constants.
The final solution is expressed in terms of the eccentric anomaly and the unperturbed (conservative) values of the semilatus rectum and radial eccentricity.
We will write below the solution for the radiation-reacted orbit (truncated at $O(G^3)$ for simplicity) in terms of the rescaled time variable $T=\frac{ p_\infty t}{b }$ and the gauge-invariant parameters $p_\infty$ and $b$.

We can distinguish a conservative part and a rr part
\bea
\label{rrorbharm}
r_h(t)&=&r_h^{\rm cons}(t)+\delta^{\rm rr}r_h(t)
\,,\nonumber\\
\phi_h(t)&=&\phi_h^{\rm cons}(t)+\delta^{\rm rr}\phi_h(t)
\,,
\eea
with
\bea
\label{deltarrorbharm}
\delta^{\rm rr}r_h(t)&=&\eta^5 \delta_{\rm 2.5PN}^{\rm rr}r_h(t)+\eta^7 \delta_{\rm 3.5PN}^{\rm rr}r_h(t)
\,,\nonumber\\
\delta^{\rm rr}\phi_h(t) &=&\eta^5 \delta_{\rm 2.5PN}^{\rm rr}\phi_h(t)+\eta^7 \delta_{\rm 3.5PN}^{\rm rr}\phi_h(t)\,.
\eea
Each PN coefficient can also be PM-expanded in powers of $G$, or equivalently for large values of the impact parameter, so that one can also write\footnote{
Note that when working in Cartesian coordinates we have $x^i=x^i_{\rm cons}(t)+\delta^{\rm rr}x^i(t)$ with
\bea
\delta^{\rm rr}x^i &=& \frac{G^2 M^2 \nu p_\infty}{b} \delta^{\rm rr,G^2} x^i +\frac{G^3 M^3 \nu }{p_\infty b^2} \delta^{\rm rr,G^3} x^i\,.\nonumber
\eea
Consequently, the relation with the polar representation of the orbits $r(t)=r_{\rm cons}(t)+\delta^{\rm rr}r(t) $ and $\phi(t)=\phi_{\rm cons}(t)+\delta^{\rm rr}\phi(t) $ is also equivalent to
\bea
\delta^{\rm rr}r(t) &=&{\mathbf n}_{\rm cons}(t)\cdot {\mathbf x}_{\rm rr}(t) \nonumber\\
\delta^{\rm rr}\phi(t) 
&=& \frac{1} {r_{\rm cons}(t)}[{\mathbf n}_{\rm cons}(t)\times {\mathbf x}_{\rm rr}(t) ]^z \nonumber\,.
\eea
}
\bea
\delta^{\rm rr}r_h(t) &=& \frac{G^2 M^2 \nu p_\infty}{b} \delta^{\rm rr,G^2} r_h(t)\nonumber\\ 
&+& \frac{G^3 M^3 \nu }{p_\infty b^2} \delta^{\rm rr,G^3} r_h(t)+O(G^4)
\,,\nonumber\\
\delta^{\rm rr}\phi_h(t) &=& \frac{G^2 M^2 \nu p_\infty}{b^2} \delta^{\rm rr,G^2} \phi_h(t)\nonumber\\
& +&\frac{G^3 M^3 \nu }{p_\infty b^3} \delta^{\rm rr,G^3}\phi_h(t)+O(G^4)\,,
\eea
where 
\begin{widetext}
\bea
\delta^{\rm rr,G^2} r_h(t) &=& -\frac{8}{5 \sqrt{T^2+1}}
+\left[
\frac{\frac{8 \nu }{5}-\frac{116}{35}}{\sqrt{T^2+1}}+\frac{\frac{4}{5}-\frac{8 \nu }{5}}{\left(T^2+1\right)^{3/2}}+T \left(\frac{8 (\nu -4)}{5 \left(T^2+1\right)}-\frac{8}{3
   \left(T^2+1\right)^2}\right)
\right]p_\infty^2
+O(p_\infty^4)
\,,\nonumber\\
\delta^{\rm rr,G^3} r_h(t) &=& {\rm at}(T) \left(-\frac{37}{15} \sqrt{T^2+1}-\frac{2}{3 \sqrt{T^2+1}}\right)-\frac{8}{5 \left(T^2+1\right)}+T \left(\frac{8 {\rm ash}(T)}{5
   \left(T^2+1\right)^{3/2}}-\frac{13}{15 \sqrt{T^2+1}}\right)+\frac{8}{5}\nonumber\\
&+&
\left[
{\rm at}(T) \left(\left(\frac{143}{168}-\frac{37 \nu }{5}\right) \sqrt{T^2+1}+\frac{\frac{121 \nu
   }{15}-\frac{107}{15}}{\sqrt{T^2+1}}+\frac{\frac{1}{3}-\frac{2 \nu }{3}}{\left(T^2+1\right)^{3/2}}\right)
+\frac{\frac{12}{5}-\frac{24 \nu
   }{5}}{\left(T^2+1\right)^2}+\frac{\frac{12 \nu }{5}-\frac{12}{7}}{T^2+1}\right.\nonumber\\
&+&
\left(\frac{\frac{32}{5}-\frac{8 \nu }{5}}{T^2+1}+\frac{\frac{16 \nu
   }{5}-\frac{24}{5}}{\left(T^2+1\right)^2}-\frac{32}{3 \left(T^2+1\right)^3}\right) {\rm ash}(T)
+\frac{12 \nu }{5}-\frac{24}{35}\nonumber\\
&+&\left.
T \left(\frac{\frac{139}{840}-5 \nu
   }{\sqrt{T^2+1}}+\frac{\frac{11 \nu }{3}-\frac{265}{12}}{\left(T^2+1\right)^{3/2}}+\left(\frac{\frac{32}{35}-\frac{4 \nu
   }{5}}{\left(T^2+1\right)^{3/2}}+\frac{\frac{24 \nu }{5}-\frac{12}{5}}{\left(T^2+1\right)^{5/2}}\right) {\rm ash}(T)-\frac{32}{3
   \left(T^2+1\right)^{5/2}}\right)
\right]p_\infty^2\nonumber\\
&+&
O(p_\infty^4)
\,,
\eea
with ${\rm at}(T)=\arctan(T)+\frac{\pi}{2}$ and ${\rm ash}(T)={\rm arcsinh}(T)$, and
\bea
\delta^{\rm rr,G^2} \phi_h(t) &=& \frac{8 T}{5 \left(T^2+1\right)}+\frac{8}{5 \sqrt{T^2+1}}
+\left[
\frac{\frac{116}{35}-\frac{8 \nu }{5}}{\sqrt{T^2+1}}+\frac{\frac{12 \nu }{5}-\frac{62}{15}}{\left(T^2+1\right)^{3/2}}-\frac{88}{15
   \left(T^2+1\right)^{5/2}}\right.\nonumber\\
&+&\left.
T \left(\frac{348-168 \nu }{105 \left(T^2+1\right)}+\frac{\frac{16 \nu
   }{5}-\frac{8}{5}}{\left(T^2+1\right)^2}\right)
\right]p_\infty^2\
+O(p_\infty^4)
\,,\nonumber\\
\delta^{\rm rr,G^3} \phi_h(t) &=& \frac{2 {\rm at}(T) T}{3 \left(T^2+1\right)}+\frac{46}{15 \left(T^2+1\right)}+T \left(\frac{8}{5 \sqrt{T^2+1}}+\frac{16}{5
   \left(T^2+1\right)^{3/2}}-\frac{8 {\rm ash}(T)}{5 \left(T^2+1\right)^{3/2}}\right)+\frac{8}{5}\nonumber\\
&+&
\left(\frac{16}{5 \left(T^2+1\right)^2}-\frac{8}{5
   \left(T^2+1\right)}\right) {\rm ash}(T)
+\left[
{\rm at}(T) T \left(\frac{\frac{59}{10}-\frac{28 \nu }{5}}{T^2+1}+\frac{\frac{4 \nu
   }{3}-\frac{2}{3}}{\left(T^2+1\right)^2}\right)+\frac{\frac{789}{70}-\frac{46 \nu }{5}}{T^2+1}+\frac{\frac{69 \nu
   }{5}-\frac{173}{10}}{\left(T^2+1\right)^2}\right.\nonumber\\
&+&
\left(\frac{\frac{232}{35}-\frac{56 \nu }{5}}{\left(T^2+1\right)^2}+\frac{\frac{4 \nu
   }{5}-\frac{32}{35}}{T^2+1}+\frac{\frac{64 \nu }{5}-\frac{32}{5}}{\left(T^2+1\right)^3}\right) {\rm ash}(T)
+\frac{4 \nu }{5}-\frac{88}{3
   \left(T^2+1\right)^3}+\frac{40}{7}\nonumber\\
&+&\left.
T\left(\frac{\frac{288}{35}-\frac{16 \nu }{5}}{\left(T^2+1\right)^{3/2}}+\frac{\frac{4 \nu }{5}+\frac{40}{7}}{\sqrt{T^2+1}}+\frac{\frac{64 \nu
   }{5}-\frac{32}{5}}{\left(T^2+1\right)^{5/2}}+\left(\frac{\frac{62}{5}-\frac{36 \nu }{5}}{\left(T^2+1\right)^{5/2}}+\frac{\frac{4 \nu
   }{5}-\frac{32}{35}}{\left(T^2+1\right)^{3/2}}+\frac{88}{3 \left(T^2+1\right)^{7/2}}\right) {\rm ash}(T)\right)
\right]p_\infty^2\nonumber\\
&+&
O(p_\infty^4)
\,.
\eea
\end{widetext}

One can easily verify that the above solution satisfies the equations of motion with the harmonic components \eqref{Frphiharm} of the rr force.
In order to get the corresponding solution for the radiation-reacted orbit in a different (Hamiltonian-based) coordinate system it is enough to invert the transformations \eqref{r_h_phi_h}, i.e.,  
\bea
r&=&r_h+O(\eta^2)\,,\qquad \phi=\phi_h+O(\eta^2)
\,,\nonumber\\
p_r&=&\dot r_h+O(\eta^2)\,,\qquad L=r_h^2\dot\phi_h+O(\eta^2)\,,
\eea
up to $O(\eta^7)$, substituting then the harmonic coordinate solution \eqref{rrorbharm}.
For a generic coordinate system $X$ we will write the rr correction to the orbit in the form
\bea
\label{deltarrorbX}
\delta^{\rm rr}r_X(t)&=&\eta^5 \delta_{\rm 2.5PN}^{\rm rr}r_X(t)+\eta^7 \delta_{\rm 3.5PN}^{\rm rr}r_X(t)
\,,\nonumber\\
\delta^{\rm rr}\phi_X(t) &=&\eta^5 \delta_{\rm 2.5PN}^{\rm rr}\phi_X(t)+\eta^7 \delta_{\rm 3.5PN}^{\rm rr}\phi_X(t)\,.
\eea
To compute the 3.5PN coefficients we also need the conservative orbit up to the 1PN accuracy, i.e.,
\bea
r_h^{\rm cons}(t)&=&r_h^{\rm N}(t)+\eta^2 r_h^{\rm 1PN}(t)+O(\eta^4)
\,,\nonumber\\
\phi_h^{\rm cons}(t) &=&\phi_h^{\rm N}(t)+\eta^2 \phi_h^{\rm 1PN}(t)+O(\eta^4)\,,
\eea
with 
\begin{widetext}
\bea
r_h^{\rm N}(t)&=&b\sqrt{1+T^2}
+\frac{1}{p_\infty^2}\left(\frac{T{\rm ash}(T)}{\sqrt{T^2+1}}-1\right)G
+O(G^2)
\,,\nonumber\\
r_h^{\rm 1PN}(t)&=&bp_\infty^2\frac{(2 \nu -1) T^2}{2 \sqrt{T^2+1}}
+\left[\frac{\frac{1}{2}-\nu }{T^2+1}+T \left(\frac{3 \left(\nu -\frac{4}{3}\right)}{2 \sqrt{T^2+1}}+\frac{\nu
   -\frac{1}{2}}{\left(T^2+1\right)^{3/2}}\right){\rm ash}(T)+\nu -\frac{5}{2}\right]G
+O(G^2)
\,,
\eea
and
\bea
\phi_h^{\rm N}(t)&=&\arctan(T)
+\frac{1}{bp_\infty^2}\left(\frac{T}{\sqrt{T^2+1}}+\frac{{\rm ash}(T)}{T^2+1}\right)G
+O(G^2)
\,,\nonumber\\
\phi_h^{\rm 1PN}(t)&=&p_\infty^2\frac{(2 \nu -1) T}{2(1+T^2)}
+\frac{1}{b}\left[T \left(\frac{\nu +4}{2\sqrt{T^2+1}}+\frac{4 \nu -2}{2\left(T^2+1\right)^{3/2}}\right)+{\rm ash}(T)\left(\frac{4 \nu -2}{2\left(T^2+1\right)^2}-\frac{\nu +2}{2\left(T^2+1\right)}\right)\right]G
+O(G^2)
\,.\nonumber\\
\eea

\subsection{ADM}

The rr correction to the orbit in ADM coordinates is given by Eq. \eqref{deltarrorbX} with
\bea
\delta_{\rm 2.5PN}^{\rm rr}r_{\rm ADM}(t)&=& \delta_{\rm 2.5PN}^{\rm rr}r_h(t) -\frac{8}{15}\nu \frac{\dot r_h^{\rm N}}{r_h^{\rm N}} 
\,,\nonumber\\
\delta_{\rm 3.5PN}^{\rm rr}r_{\rm ADM}(t)&=& \delta_{\rm 3.5PN}^{\rm rr}r_h(t)
-\frac{8}{15}\nu\frac{1}{(r_h^{\rm N})^2}(-r_h^{\rm 1PN}\dot r_h^{\rm N}+\dot r_h^{\rm 1PN}r_h^{\rm N})\nonumber\\
&+& 
\nu\frac{\dot r_h^{\rm N}}{r_h^{\rm N}}\left[
\left(\frac{116\nu }{105}+\frac{292}{105}\right)(v_h^{\rm N})^2
+\left(-\frac{148 \nu }{105}-\frac{176}{105}\right)(\dot r_h^{\rm N})^2
+\left(\frac{44 \nu }{35}+\frac{125}{21}\right)\frac{1}{r_h^{\rm N}}
\right]
\,,
\eea
where $(v_h^{\rm N})^2=(\dot r_h^{\rm N})^2+(r_h^{\rm N})^2(\dot \phi_h^{\rm N})^2$, and
\bea
\delta_{\rm 2.5PN}^{\rm rr}\phi_{\rm ADM}(t)&=& \delta_{\rm 2.5PN}^{\rm rr}\phi_{\rm h}(t) +\frac{16}{15}\nu \frac{\dot \phi_h^{\rm N}}{r_h^{\rm N}}     
\,,\nonumber\\
\delta_{\rm 3.5PN}^{\rm rr}\phi_{\rm ADM}(t)&=& \delta_{\rm 3.5PN}^{\rm rr}\phi_{\rm h}(t)
+\frac{16}{15}\nu\frac{1}{(r_h^{\rm N})^2}(-r_h^{\rm 1PN}\dot\phi_h^{\rm N}+\dot\phi_h^{\rm 1PN}r_h^{\rm N})\nonumber\\
&+& 
\nu\frac{\dot\phi_h^{\rm N}}{r_h^{\rm N}}\left[
\left(\frac{12 \nu}{35}+\frac{236}{105}\right)(v_h^{\rm N})^2
+\left(-\frac{44 \nu}{21}-\frac{374}{105}\right)(\dot r_h^{\rm N})^2
+\left(-\frac{92 \nu}{35}-\frac{424}{105}\right)\frac{1}{r_h^{\rm N}}
\right]
\,.
\eea

\subsection{EOB}

The rr correction to the orbit in EOB coordinates is given by Eq. \eqref{deltarrorbX} with
\bea
\delta_{\rm 2.5PN}^{\rm rr}r_{\rm EOB}(t)&=& \delta_{\rm 2.5PN}^{\rm rr}r_h(t) + \frac{8}{5}\frac{\dot r_h^{\rm N}}{r_h^{\rm N}}\nu (\alpha-\beta+1) 
\,,\nonumber\\
\delta_{\rm 3.5PN}^{\rm rr}r_{\rm EOB}(t)&=& \delta_{\rm 3.5PN}^{\rm rr}r_h(t)
-\nu(r_h^{\rm N})^3(\dot\phi_h^{\rm N})\delta_{\rm 2.5PN}^{\rm rr}\dot \phi_{\rm h}(t)  
-3\nu(r_h^{\rm N})(\dot r_h^{\rm N})\delta_{\rm 2.5PN}^{\rm rr}\dot r_h(t) 
-\frac32\nu(v_h^{\rm N})^2\delta_{\rm 2.5PN}^{\rm rr}r_h(t)\nonumber\\
&+& 
\frac{8}{5}\nu\frac{\alpha-\beta+1}{(r_h^{\rm N})^2}(-r_h^{\rm 1PN}\dot r_h^{\rm N}+\dot r_h^{\rm 1PN}r_h^{\rm N})\nonumber\\
&+& 
\nu\dot r_h^{\rm N}\left[
 \left(\nu\Big(\tfrac{14}{15}\alpha-\tfrac{23}{30}\beta+\tfrac{334}{105}\Big)+\tfrac2{15}\beta-\tfrac25\delta_1-\tfrac13\delta_2+\tfrac15\delta_3-\tfrac13\delta_4+\tfrac1{10}\delta_5-\tfrac1{15}\epsilon_5-\tfrac{587}{420} \right)(v_h^{\rm N})^4\right.\nonumber\\
&+& 
 \left(\nu\Big(-\tfrac{24}{5}\alpha+\tfrac{46}{15}\beta-\tfrac{596}{35}\Big)-\tfrac{12}5\alpha+\tfrac43\beta-\tfrac4{15}\delta_2-\tfrac45\delta_3+\tfrac4{15}\delta_4-\tfrac25\delta_5+\tfrac45\epsilon_5+\tfrac{145}{21} \right)\frac{(v_h^{\rm N})^2}{r_h^{\rm N}}\nonumber\\
&+&\left( \nu\Big(\tfrac{16}{15}\alpha+\tfrac{40}{21}\Big)+\tfrac8{15}\beta-\tfrac8{15}\delta_4-\tfrac8{15}\epsilon_5-\tfrac{106}{35}\right) 
 \frac{(\dot r_h^{\rm N})^2}{r_h^{\rm N}}\nonumber\\
&+&\left. 
\left( \nu\Big(\tfrac{32}{15}\alpha-\tfrac{22}{15}\beta+\tfrac{204}{35}\Big)+\tfrac8{15}\beta+\tfrac4{15}\delta_2+\tfrac4{15}\delta_4-\tfrac25\delta_5-\tfrac4{15}\epsilon_5-\tfrac53\right)
 \frac{1}{(r_h^{\rm N})^2}
\right]
\,,
\eea
and
\bea
\delta_{\rm 2.5PN}^{\rm rr}\phi_{\rm EOB}(t)&=& \delta_{\rm 2.5PN}^{\rm rr}\phi_{\rm h}(t)
+\frac{8}{15}\nu(3\alpha-2\beta+3)\frac{\dot \phi_h^{\rm N}}{r_h^{\rm N}} 
\,,\nonumber\\
\delta_{\rm 3.5PN}^{\rm rr}\phi_{\rm EOB}(t)&=& \delta_{\rm 3.5PN}^{\rm rr}\phi_{\rm h}(t)  
-\nu\dot\phi_h^{\rm N}(\dot r_h^{\rm N}\delta_{\rm 2.5PN}^{\rm rr}r_h(t) + r_h^{\rm N}\delta_{\rm 2.5PN}^{\rm rr}\dot r_h(t))
-\nu r_h^{\rm N}\dot r_h^{\rm N}\delta_{\rm 2.5PN}^{\rm rr}\dot \phi_{\rm h}(t) \nonumber\\
&+&   
\frac{8}{15}\nu\frac{3\alpha-2\beta+3}{(r_h^{\rm N})^2}(-r_h^{\rm 1PN}\dot\phi_h^{\rm N}+\dot\phi_h^{\rm 1PN}r_h^{\rm N})\nonumber\\
&+& 
\nu\dot \phi_h^{\rm N}\left[
\left(\nu\Big(\tfrac{14}{15}\alpha-\tfrac{23}{30}\beta+\tfrac{334}{105}\Big)+\tfrac2{15}\beta-\tfrac25\delta_1-\tfrac13\delta_2+\tfrac15\delta_3-\tfrac13\delta_4+\tfrac1{10}\delta_5-\tfrac1{15}\epsilon_5-\tfrac{587}{420} \right)
 (v_h^{\rm N})^4\right.\nonumber\\
&+& \left( \nu\Big(-\tfrac{244}{75}\alpha+\tfrac{38}{15}\beta-\tfrac{5756}{525}\Big)-\tfrac{12}5\alpha+\tfrac{16}{15}\beta+\tfrac4{15}\delta_2-\tfrac45\delta_3+\tfrac{12}{25}\delta_4-\tfrac25\delta_5+\tfrac4{15}\epsilon_5+\tfrac{701}{105}\right)
 \frac{(v_h^{\rm N})^2}{r_h^{\rm N}}\nonumber\\
&+&\left.
\left( \nu\Big(\tfrac{224}{75}\alpha-\tfrac85\beta+\tfrac{2176}{525}\Big)+\tfrac8{15}\beta-\tfrac{16}{75}\delta_4-\tfrac8{15}\epsilon_5-\tfrac{218}{35}\right) 
\frac{(\dot r_h^{\rm N})^2}{r_h^{\rm N}}
+\frac{4}{15} \nu  (3 \alpha -4 \beta+3)
\frac{1}{(r_h^{\rm N})^2}
\right]
\,.\nonumber\\
\eea

\end{widetext}

\section{Concluding remarks}

We have considered the relative dynamics of a gravitational two-body system up to the 3.5PN order, including both conservative contributions and radiation-reaction effects (starting at the 2.5PN level).
The full computation of the radiation-reaction force would require a direct integration of the field equations in the source's near zone.
This is a hard task even at the lowest PN orders, and has been indeed done in few cases only: harmonic, BT, and ADM coordinates, for which the radiation-reaction force has been fully determined in a general frame.
In some cases such a first-principle approach cannot be adopted, as in the EOB framework. 

A less expensive method (even if restricted to the cm frame) consists in assuming a balance equation between the instantaneous (coordinate-dependent) energy and angular momentum losses in the near zone and the (gauge-invariant) instantaneous energy and angular momentum fluxes in the far zone.
However, in this way there is an ambiguity in the choice of the gauge at each PN order, since the balance actually holds up to total time derivatives, corresponding to Schott terms.
As a result, the rr force is not uniquely determined by the balance equations, but contains several arbitrary parameters representing the gauge freedom. 
Without additional information this freedom can be used to simplify the expressions for the components of the rr force, but this is more a choice than a rule. 
This is the case of the EOB rr force, for which several expressions have been used in the literature to construct waveform models corresponding to different (often not well motivated) choices of the gauge parameters.

Starting from the harmonic-coordinate expressions for the losses of mechanical energy and angular momentum we have provided a general prescription to obtain the radial and azimuthal components of the radiation-reaction force in a different coordinate system expressed in terms of phase-space variables in a Hamiltonian framework.
There is no need to solve again the balance equations, only the coordinate transformation between harmonic coordinates and coordinates and momenta in the new coordinate system is required.
We have explicitly derived such a transformation for both ADM and EOB coordinate systems.
We have also shown how to get the solution for the radiation-reaction correction to the orbit in the new coordinate system simply by inverting such a transformation, and using the harmonic-coordinate solution known in the literature. 
This is an important result of the present work if one considers that for any chosen coordinate system one should apply the Lagrange method of variation of arbitrary constants, leading to a nontrivial system of coupled differential equations for the time variation of the orbital parameters.

Our approach can be straightforwardly generalized to higher PN orders by including nonlocal effects too. The harmonic-coordinate rr force has been indeed recently derived in the spinless case at the 4.5PN order. More importantly, it is also well suited to incorporating spin corrections, which currently represent one of the main challenges in the  gravitational wave  research area. We plan to pursue both directions in future works.

\section*{Acknowledgments}

The authors are indebted to  T. Damour for a careful reading of a preliminary version of the manuscript and useful comments and suggestions.
They also acknowledge informative discussions with L. Blanchet and G. Faye.
D.B. and S.R.A acknowledge membership to the Italian Gruppo Nazionale per la Fisica Matematica (GNFM) of the Istituto Nazionale
di Alta Matematica (INDAM).
A.G.  is grateful to the Istituto per le Applicazioni del Calcolo \lq\lq M. Picone," CNR, Rome (IT) for past support and hospitality during the development of the present project.

\appendix

\section{Maps from harmonic coordinates to ADM and EOB phase-space coordinates}

We list below the transformations from harmonic coordinates to ADM and EOB phase-space coordinates.
We will set $G=1=M$ for the sake of simplicity. We also recall the notation $p_\phi=L$.

\begin{widetext}
\subsection{Harmonic to ADM}

The transformation between harmonic coordinates and ADM coordinates is known for the conservative part up to the 3PN order (see Ref. \cite{Blanchet:2002mb}).
The rr part is new with this work.
We find
\bea
r_h &=& r+\eta^4\left[-\frac{7}{4}\nu \frac{p_r^2}{r}+\frac{5}{8}\frac{L^2}{r^3}\nu+\left(\frac14+3\nu\right)\frac{1}{r^2}\right]r+\frac{8}{15}\eta^5 \nu \frac{p_r}{r}\nonumber\\
&+&\eta^6 \left[\left(-\frac{47}{24}\nu^2+\frac{7}{24}\nu \right) \frac{p_r^4}{r}
+\left(\frac{1}{16}\nu^2-\frac{5}{16}\nu \right) p_r^2\frac{L^2}{r^3}+\left(\frac{39}{8}\nu^2+\frac{53}{24}\nu \right) \frac{p_r^2}{r^2}
+\left(-\frac18 \nu+\frac12 \nu^2\right) \frac{L^4}{r^5}\right.\nonumber\\
&+&\left. \left(-\frac78 \nu^2+\frac{271}{48}\nu \right) \frac{L^2}{r^4}+\left(-\frac{2773}{280}\nu-\frac{21}{32}\nu\pi^2\right)\frac{1}{r^3}\right] r\nonumber\\
&+&\eta^7 \left[\left(-\frac{48}{35}\nu+\frac{116}{105}\nu^2\right)\frac{ p_r^2}{r^2}+\left(-\frac{64}{21}\nu-\frac{32}{105}\nu^2\right) \frac{L^2}{r^4}
+\left(-\frac{793}{105}\nu-\frac{244}{105}\nu^2 \right)\frac{1}{r^3}\right] p_r r
\,,\nonumber\\
%%%
\phi_h &=& \phi-\frac94 \eta^4 \nu \frac{p_r L}{r^2}-\frac{16}{15}\eta^5 \nu \frac{L}{r^3}\nonumber\\
&+&\eta^6\left[\left(-\frac{65}{24}\nu^2+\frac{13}{24}\nu \right) \frac{p_r^2}{r}
+\left(-\frac32 \nu^2+\frac18 \nu \right)\frac{L^2}{r^3}
+\left(\frac74\nu^2-\frac56\nu \right)\frac{1}{r^2}\right] p_r \frac{L}{r}\nonumber\\
&+&\eta^7\left[\left(\frac{194}{105}\nu+\frac{16}{105}\nu^2 \right) \frac{p_r^2}{r^2}+\left(-\frac{12}{7}\nu-\frac{68}{35}\nu^2 \right)\frac{L^2}{r^4}+\left(\frac{388}{105}\nu^2+\frac{152}{21}\nu \right)\frac{1}{r^3}\right]\frac{L}{r}\,,
\eea
and
\bea
\dot r_h &=& p_r
+\eta ^2 \left[\frac{L^2 p_r}{r^2}
   \left(\frac{3 \nu }{2}-\frac{1}{2}\right)+\left(\frac{3 \nu }{2}-\frac{1}{2}\right) p_r^3+(-2 \nu -3)\frac{p_r}{r}\right]\nonumber\\
&+&
\eta ^4 \left[\frac{L^4p_r}{r^4} \left(\frac{15 \nu ^2}{8}-\frac{15 \nu }{8}+\frac{3}{8}\right) +
   \frac{L^2p_r^3}{r^2}\left(\frac{15 \nu ^2}{4}-\frac{15 \nu }{4}+\frac{3}{4}\right) +\frac{L^2p_r}{r^3}\left(-2 \nu ^2-\frac{59 \nu}{4}+\frac{5}{2}\right)\right.\nonumber\\
&+&\left.
\left(\frac{15 \nu ^2}{8}-\frac{15 \nu }{8}+\frac{3}{8}\right) p_r^5+\frac{p_r^3}{r}\left(-4 \nu
   ^2-10 \nu +\frac{5}{2}\right) +\frac{p_r}{r^2}\left(\frac{23 \nu }{2}+\frac{19}{4}\right) \right]
+\eta ^5 \frac{8 \nu }{15 r^2}\left(\frac{L^2}{r^2}-p_r^2-\frac{1}{r}\right)\nonumber\\
&+&
\eta ^6 \left[\frac{L^6p_r}{r^6} \left(\frac{35 \nu ^3}{16}-\frac{35 \nu ^2}{8}+\frac{35 \nu
   }{16}-\frac{5}{16}\right)
+\frac{L^4p_r^3}{r^4} \left(\frac{105 \nu ^3}{16}-\frac{105 \nu ^2}{8}+\frac{105 \nu
   }{16}-\frac{15}{16}\right)\right.\nonumber\\
&+& 
\frac{L^4p_r}{r^5} \left(-\frac{9 \nu ^3}{4}-\frac{229 \nu ^2}{8}+18 \nu -\frac{21}{8}\right)
+\frac{L^2p_r}{r^4} \left(\frac{93 \nu ^2}{2}+\frac{2153 \nu }{48}-\frac{53}{8}\right) 
+\frac{L^2p_r^5}{r^2} \left(\frac{105\nu ^3}{16}-\frac{105 \nu ^2}{8}+\frac{105 \nu }{16}-\frac{15}{16}\right)\nonumber\\
&+& 
\frac{L^2p_r^3}{r^3} \left(-6 \nu ^3-\frac{637 \nu^2}{12}+\frac{107 \nu }{3}-\frac{21}{4}\right)
+\frac{p_r^3}{r^2}\left(\frac{1225 \nu ^2}{24}+\frac{515 \nu}{12}-\frac{53}{8}\right) 
+\frac{p_r}{r^3}\left(-13 \nu ^2+\frac{5 \pi ^2 \nu }{4}-\frac{447 \nu }{140}-\frac{11}{2}\right)\nonumber\\
&+&\left.
\left(\frac{35 \nu ^3}{16}-\frac{35 \nu ^2}{8}+\frac{35 \nu }{16}-\frac{5}{16}\right) p_r^7
+\frac{p_r^5}{r}\left(-6 \nu ^3-18 \nu^2+\frac{63 \nu }{4}-\frac{21}{8}\right) \right]\nonumber\\
&+&
\eta ^7 \left[\frac{L^4}{r^6} \left(\frac{52 \nu ^2}{105}-\frac{116 \nu }{35}\right)
+\frac{L^2p_r^2}{r^4}\left(\frac{148 \nu ^2}{35}+\frac{176 \nu}{35}\right) 
+\frac{L^2}{r^5}\left(-\frac{296 \nu ^2}{105}-\frac{145 \nu }{21}\right)
+\frac{p_r^4}{r^2}\left(\frac{172 \nu }{105}-\frac{40 \nu^2}{21}\right)\right.\nonumber\\
&+&\left. 
\frac{p_r^2}{r^3}\left(\frac{28 \nu ^2}{15}+\frac{2102 \nu }{105}\right) 
+\frac{1}{r^4}\left(\frac{244 \nu^2}{105}+\frac{283 \nu }{35}\right)\right]
\,,\nonumber\\
%%%
\dot\phi_h &=&\frac{L}{r^2}
+\eta ^2 \left[\frac{L^3}{r^4} \left(\frac{3 \nu }{2}-\frac{1}{2}\right)
+\frac{Lp_r^2}{r^2}\left(\frac{3 \nu}{2}-\frac{1}{2}\right) 
+\frac{L}{r^3}\left(-\nu -3\right)\right]\nonumber\\
&+&
\eta ^4 \left[\frac{L^5}{r^6} \left(\frac{15 \nu^2}{8}-\frac{15 \nu }{8}+\frac{3}{8}\right)
+\frac{L^3p_r^2}{r^4}\left(\frac{15 \nu ^2}{4}-\frac{15 \nu }{4}+\frac{3}{4}\right)
+\frac{L^3}{r^5}\left(-\frac{3 \nu ^2}{2}-\frac{49 \nu }{4}+\frac{5}{2}\right)\right.\nonumber\\
&+&\left.
\frac{Lp_r^4}{r^2}\left(\frac{15 \nu ^2}{8}-\frac{15 \nu}{8}+\frac{3}{8}\right) 
+\frac{Lp_r^2}{r^3}\left(-2 \nu ^2-\frac{11 \nu }{2}+\frac{5}{2}\right) 
+\frac{L}{r^4}\left(\frac{41 \nu}{4}+5\right)\right]
+ \eta ^5\frac{16 \nu  L p_r}{5 r^4}
\nonumber\\
&+&
\eta ^6
   \left[\frac{L^7}{r^8} \left(\frac{35 \nu ^3}{16}-\frac{35 \nu ^2}{8}+\frac{35 \nu }{16}-\frac{5}{16}\right)
+\frac{L^5p_r^2}{r^6}\left(\frac{105 \nu^3}{16}-\frac{105 \nu ^2}{8}+\frac{105 \nu }{16}-\frac{15}{16}\right)\right.\nonumber\\
&+&\left. 
\frac{L^5}{r^7}\left(-\frac{15 \nu ^3}{8}-\frac{99 \nu ^2}{4}+17 \nu -\frac{21}{8}\right)
+\frac{L^3}{r^6}\left(\frac{271 \nu ^2}{8}+\frac{259 \nu }{6}-\frac{27}{4}\right)
+\frac{L^3p_r^4}{r^4}\left(\frac{105 \nu^3}{16}-\frac{105 \nu ^2}{8}+\frac{105 \nu }{16}-\frac{15}{16}\right)\right.\nonumber\\
&+& 
\frac{L^3p_r^2}{r^5}\left(-\frac{9 \nu ^3}{2}-38 \nu ^2+\frac{63 \nu}{2}-\frac{21}{4}\right)
+\frac{Lp_r^2}{r^4}\left(\frac{217 \nu ^2}{8}+\frac{215 \nu }{8}-\frac{27}{4}\right)
+\frac{L}{r^5}\left(-\frac{15 \nu ^2}{2}+\frac{\pi ^2 \nu }{32}-\frac{123 \nu }{8}-\frac{25}{4}\right)\nonumber\\
&+&\left.
\frac{Lp_r^6}{r^2}\left(\frac{35 \nu^3}{16}-\frac{35 \nu ^2}{8}+\frac{35 \nu }{16}-\frac{5}{16}\right)
+\frac{Lp_r^4}{r^3}\left(-3 \nu ^3-\frac{41 \nu ^2}{6}+\frac{149 \nu}{12}-\frac{21}{8}\right)\right]
\nonumber\\
&+&
\eta ^7 \left[\frac{L^3p_r}{r^6} \left(\frac{1556 \nu ^2}{105}+\frac{32 \nu }{3}\right) 
+\frac{Lp_r^3}{r^4}\left(\frac{152 \nu ^2}{35}-\frac{50\nu }{7}\right) 
+\frac{Lp_r}{r^5}\left(-\frac{752 \nu ^2}{35}-\frac{4436 \nu }{105}\right)\right]
\,,
\eea

\subsection{Harmonic to EOB}

The transformation between harmonic coordinates and EOB coordinates has been derived through the 3PN order in Ref. \cite{Gamboa:2024imd} (see Appendix A there).
We extend it here to the 3.5PN level, so that besides $\alpha$ and $\beta$ it also contains the six 3.5PN gauge parameters $\delta_i$, $i=1,\ldots,5$, and $\epsilon_5$. 
We find
\bea
r_h &=&r
+\eta ^2 \left[\frac{L^2 \nu }{2 r^2}+\frac{3 \nu  p_r^2}{2}+\frac{1}{r}\left(-\frac{\nu}{2}-1\right)\right]r \nonumber\\
&+&
\eta ^4 \left[\frac{L^4}{r^4} \left(-\frac{\nu^2}{8}-\frac{\nu }{8}\right)
+\frac{L^2p_r^2}{r^2} \left(\frac{3 \nu ^2}{4}-\frac{3 \nu }{4}\right) 
+\frac{L^2}{r^3} \left(\frac{3 \nu^2}{8}-\frac{\nu }{8}\right)
+\left(\frac{3 \nu ^2}{8}-\frac{5 \nu }{8}\right) p_r^4
+\frac{1}{r^2}\left(\frac{19 \nu }{4}-\frac{\nu^2}{4}\right)-\frac{7\nu p_r^2}{r} \right]r \nonumber\\
&-&
\eta ^5\frac{8\nu (1+\alpha-\beta)p_r}{5r}\nonumber\\
&+&
\eta ^6 \left[\frac{L^6}{r^6}\left(\frac{\nu ^3}{16}+\frac{\nu ^2}{16}+\frac{\nu }{16}\right)
+\frac{L^4p_r^2}{r^4} \left(\frac{5 \nu ^3}{16}-\frac{15 \nu ^2}{16}+\frac{9 \nu}{16}\right) 
+\frac{L^4}{r^5} \left(-\frac{5 \nu ^3}{16}+\frac{\nu ^2}{16}+\frac{\nu }{16}\right)
+\frac{L^2}{r^4} \left(\frac{\nu ^3}{2}-\frac{11 \nu ^2}{4}+\frac{91 \nu }{12}\right)\right.\nonumber\\
&+&
\frac{L^2p_r^4}{r^2} \left(-\frac{5 \nu^3}{16}-\frac{25 \nu ^2}{16}+\frac{15 \nu }{16}\right) 
+\frac{L^2p_r^2}{r^3} \left(\frac{7 \nu ^3}{16}-\frac{135 \nu ^2}{16}+\frac{37 \nu}{16}\right)
+\left(-\frac{\nu^3}{16}-\frac{9 \nu ^2}{16}+\frac{7 \nu }{16}\right) p_r^6\nonumber\\
&+&\left.
\frac{p_r^4}{r}\left(\frac{5 \nu ^3}{48}-\frac{133 \nu ^2}{48}+\frac{179 \nu}{48}\right)
+\frac{p_r^2}{r^2}\left(-\frac{\nu ^3}{8}+\frac{37 \nu ^2}{8}+\frac{275 \nu }{24}\right) 
+\frac{1}{r^3}\left(-\frac{\nu^3}{4}+\frac{7 \nu ^2}{4}-\frac{41 \pi ^2 \nu }{64}+\frac{137 \nu }{105}\right)\right]r \nonumber\\
&+&
\eta^7\left[C_1\left(p_r^2+\frac{L^2}{r^2}\right)^2\frac{1}{r}+C_2\frac{L^2}{r^4}+C_3\frac{p_r^2}{r^2}+C_4\frac{1}{r^3}\right]\nu p_r r
\,,\nonumber\\
%%%%%%%
\phi_h &=& \phi
+\eta ^2\frac{ L \nu  p_r}{r}
+\eta ^4 \left[-\frac{L^2 \nu }{2 r^2}+\left(-\nu ^2-\frac{\nu
   }{2}\right) p_r^2+\left(\frac{3 \nu ^2}{4}-\frac{15 \nu }{4}\right)\frac{1}{r}\right]\frac{L p_r}{r} 
-\eta ^5\frac{8\nu (3+3\alpha-2\beta)L}{15r^3}\nonumber\\
&+&
\eta ^6\left[\frac{L^4}{r^4} \left(\frac{3 \nu }{8}-\frac{\nu ^2}{4}\right)
+\frac{L^2p_r^2}{r^2} \left(-\frac{\nu ^3}{3}+\frac{\nu ^2}{2}+\frac{3 \nu }{4}\right)
+\frac{L^2}{r^3} \left(-\frac{\nu ^3}{8}-\frac{15 \nu ^2}{8}+\frac{7 \nu }{8}\right)
+\left(\nu ^3+\frac{3 \nu ^2}{4}+\frac{3\nu }{8}\right) p_r^4\right.\nonumber\\
&+&\left.
\frac{p_r^2}{r}\left(-\frac{19 \nu ^3}{12}+\frac{55 \nu ^2}{6}+\frac{55 \nu }{24}\right) 
+\frac{1}{r^2}\left(\frac{3 \nu^3}{4}-\frac{3 \nu ^2}{4}-\frac{95 \nu }{24}\right)\right]\frac{L p_r}{r}\nonumber\\
&+&
\eta^7\left[C_1\left(p_r^2+\frac{L^2}{r^2}\right)^2\frac{1}{r}+C_5\frac{L^2}{r^4}+C_6\frac{p_r^2}{r^2}\right]\nu\frac{L}{r}  
\,,
\eea
where
\bea
C_1&=&-\frac{14}{15}\alpha\nu+\frac{23}{30}\beta\nu-\frac{2}{15}\beta+\frac25\delta_1+\frac13\delta_2-\frac15\delta_3+\frac13\delta_4-\frac1{10}\delta_5+\frac1{15}\epsilon_5-\frac{334}{105}\nu+\frac{587}{420}\,,\nonumber\\
C_2&=&\frac{16}{5}\alpha+\frac{10}{3}\beta\nu-\frac{32}{15}\beta+\frac{4}{15}\delta_2+\frac45\delta_3-\frac4{15}\delta_4+\frac25\delta_5-\frac45\epsilon_5+\frac{428}{35}\nu-\frac{641}{105}\,,\nonumber\\
C_3&=&\frac{104}{15}\alpha\nu+\frac{16}{5}\alpha-\frac{94}{15}\beta\nu-\frac83\beta+\frac4{15}\delta_2+\frac45\delta_3+\frac4{15}\delta_4+\frac25\delta_5-\frac4{15}\epsilon_5+\frac{1924}{105}\nu-\frac{323}{105}\,,\nonumber\\
C_4&=&\frac{76}{15}\alpha\nu+\frac{16}{5}\alpha-\frac{86}{15}\beta\nu-\frac{56}{15}\beta-\frac4{15}\delta_2-\frac4{15}\delta_4+\frac25\delta_5+\frac4{15}\epsilon_5+\frac{48}{35}\nu+\frac{73}{15}\,,\nonumber\\
C_5&=&\frac{124}{75}\alpha\nu+\frac{16}{5}\alpha-\frac{14}{15}\beta\nu-\frac85\beta-\frac4{15}\delta_2+\frac45\delta_3-\frac{12}{25}\delta_4+\frac25\delta_5-\frac4{15}\epsilon_5+\frac{4916}{525}\nu-\frac{617}{105}\,,\nonumber\\
C_6&=&\frac{148}{15}\alpha\nu+\frac{16}{5}\alpha-\frac{22}{3}\beta\nu-\frac{32}{15}\beta-\frac4{15}\delta_2+\frac45\delta_3-\frac4{15}\delta_4+\frac25\delta_5+\frac4{15}\epsilon_5+\frac{1724}{105}\nu+\frac{37}{105}\,.
\eea
and
\bea
\dot r_h &=&p_r
+\eta ^2 \left[\frac{L^2p_r}{r^2} \left(2 \nu -\frac{1}{2}\right) +\left(\nu -\frac{1}{2}\right)
   p_r^3+\frac{p_r}{r}(-2 \nu -3) \right]
+\eta ^4 \left[\frac{L^4p_r}{r^4}\left(\nu ^2-2 \nu +\frac{3}{8}\right) 
+\frac{L^2p_r^3}{r^2} \left(-\frac{\nu ^2}{2}-3 \nu +\frac{3}{4}\right)\right.\nonumber\\
&+&\left.
\frac{L^2p_r}{r^3} \left(\frac{\nu ^2}{4}-\frac{55 \nu }{4}+\frac{3}{2}\right)
+\frac{p_r}{r^2}\left(-\frac{5 \nu^2}{4}+\frac{39 \nu }{4}+\frac{3}{2}\right)
+\left(\frac{3}{8}-\nu \right) p_r^5
+\frac{p_r^3}{r}\left(\frac{5}{2}-4 \nu \right) \right]\nonumber\\
&-&
\eta ^5\frac{8\nu}{5r^2}(1+\alpha-\beta) \left[\frac{L^2}{r^2}-p_r^2-\frac{1}{r}\right]
+\eta ^6 \left[\frac{L^6p_r}{r^6} \left(-2 \nu ^2+2 \nu -\frac{5}{16}\right) 
+\frac{L^4p_r^3}{r^4}\left(-2 \nu ^3-2 \nu ^2+5 \nu -\frac{15}{16}\right)\right.\nonumber\\
&+& 
\frac{L^4p_r}{r^5}\left(\frac{3 \nu ^3}{2}-\frac{103 \nu ^2}{8}+\frac{45\nu }{4}-\frac{9}{8}\right)
+\frac{L^2p_r^5}{r^2}\left(\frac{\nu ^3}{2}-\frac{\nu ^2}{4}+4 \nu -\frac{15}{16}\right)
+\frac{L^2p_r^3}{r^3}\left(-\frac{13 \nu ^3}{12}+\frac{433 \nu ^2}{24}+\frac{127 \nu }{6}-\frac{15}{4}\right)\nonumber\\
&+&
\frac{L^2p_r}{r^4}\left(-\frac{\nu ^3}{2}+\frac{119 \nu ^2}{8}+\frac{73 \nu }{6}-\frac{3}{4}\right)
+\left(-\frac{\nu ^2}{4}+\nu -\frac{5}{16}\right) p_r^7
+\frac{p_r^5}{r}\left(\frac{3 \nu ^2}{2}+6 \nu -\frac{21}{8}\right)\nonumber\\
&+&\left. 
\frac{p_r^3}{r^2}\left(\frac{13 \nu ^3}{12}+\frac{191 \nu ^2}{24}+\frac{25 \nu }{2}-\frac{15}{4}\right)
+\frac{p_r}{r^3}\left(-\nu^3-\frac{15 \nu ^2}{4}+\frac{41 \pi ^2 \nu }{32}+\frac{3751 \nu }{105}+\frac{1}{2}\right) \right]\nonumber\\
&+&\eta ^7\nu 
\Big[C_1\left(p_r^2+\frac{L^2}{r^2}\right)^2 \frac{L^2}{r^3}
+\frac{L^4}{r^6}\Big(-C_1+C_2+\frac45(\alpha-\beta+1)(1+\nu)-\frac85(\alpha+2)\nu\Big)\nonumber\\
&+&\frac{L^2p_r^2}{r^4}\Big(-3(2C_1+C_2-C_3)+\frac85(9\beta-\alpha+4)\nu\Big)
+\frac{L^2}{r^5}\Big(-C_2+C_4-\frac{12}5(\alpha-\beta+1)(\nu-1)+\frac85(\alpha-2)\nu\Big)\nonumber\\
&+&\frac{p_r^4}{r^2}\Big(-5C_1-C_3-\frac45(\alpha-\beta+1)(1+\nu)+\frac{48}5(\alpha-\beta+2)\nu\Big)\nonumber\\
&+&\frac{p_r^2}{r^3}\Big(-3C_3-2C_4+\frac45(\alpha-\beta+1)(\nu-3)+\frac85(9\alpha-9\beta+17)\nu\Big)
+\frac{1}{r^4}\Big(-C_4+\frac85(\alpha-\beta+1)(1+\nu)\Big)\Big]
\,,\nonumber\\
%%%%%
\dot\phi_h &=& \frac{L}{r^2}
+\eta ^2 \left[\frac{L^2}{r^2} \left(\frac{\nu }{2}-\frac{1}{2}\right)+\left(-\frac{3 \nu }{2}-\frac{1}{2}\right) p_r^2-\frac{1}{r}\right]\frac{L}{r^2}
+\eta ^4 \left[\frac{L^5}{r^6} \left(-\frac{\nu ^2}{8}-\frac{5 \nu }{8}+\frac{3}{8}\right)
+\frac{L^3p_r^2}{r^4}\left(-\frac{9 \nu ^2}{4}+\frac{3 \nu }{4}+\frac{3}{4}\right)\right.\nonumber\\
&+&\left. 
\frac{L^3}{r^5}\left(\frac{3 \nu^2}{4}-\frac{17 \nu }{4}+\frac{1}{2}\right)
+\frac{Lp_r^4}{r^2}\left(\frac{15 \nu ^2}{8}+\frac{11 \nu }{8}+\frac{3}{8}\right)
+\frac{Lp_r^2}{r^3}\left(-\frac{\nu ^2}{2}+12 \nu +\frac{3}{2}\right) 
+\frac{L}{r^4}\left(-\frac{\nu ^2}{4}+\frac{9 \nu}{4}-\frac{1}{2}\right)\right]\nonumber\\
&+&
\eta^5\frac{8\nu(3+3\alpha-2\beta)Lp_r}{5r^4}
+\eta ^6 \left[\frac{L^7}{r^8}\left(\frac{\nu ^3}{16}+\frac{11 \nu }{16}-\frac{5}{16}\right)
+\frac{L^5p_r^2}{r^6}\left(-\frac{\nu ^3}{16}+\frac{15 \nu ^2}{4}+\frac{\nu}{16}-\frac{15}{16}\right)\right.\nonumber\\
&+& 
\frac{L^5}{r^7}\left(-\frac{\nu ^3}{2}-\frac{\nu ^2}{4}+\frac{27 \nu }{8}-\frac{3}{8}\right)
+\frac{L^3p_r^4}{r^4}\left(\frac{91 \nu ^3}{16}+\frac{3 \nu ^2}{2}-\frac{31 \nu }{16}-\frac{15}{16}\right) 
+\frac{L^3p_r^2}{r^5}\left(-\frac{35 \nu ^3}{8}+\frac{139 \nu ^2}{4}-\frac{15 \nu }{4}-\frac{9}{4}\right)\nonumber\\
&+& 
\frac{L^3}{r^6}\left(\frac{5 \nu ^3}{4}-\frac{37 \nu ^2}{8}+\frac{13 \nu}{24}+\frac{1}{4}\right)
+\frac{Lp_r^6}{r^2}\left(-\frac{35 \nu ^3}{16}-\frac{9 \nu ^2}{4}-\frac{21 \nu }{16}-\frac{5}{16}\right)
+\frac{Lp_r^4}{r^3}\left(\frac{17 \nu ^3}{12}-\frac{653 \nu ^2}{24}-\frac{317 \nu }{24}-\frac{15}{8}\right)\nonumber\\
&+&\left.
\frac{Lp_r^2}{r^4}\left(\frac{5 \nu ^3}{8}-\frac{41 \nu ^2}{4}-\frac{113 \nu }{8}-\frac{3}{4}\right)
+\frac{L}{r^5}\left(-\frac{\nu^3}{2}+\frac{17 \nu ^2}{4}+\frac{161 \nu }{24}-\frac{1}{2}\right)\right]\nonumber\\
&+&
\eta ^7 
\frac{\nu Lp_r}{r^3}\Big[-2C_1\left(p_r^2+\frac{L^2}{r^2}\right)^2
+\frac{L^2}{r^3}\Big(-4C_1-5C_5+2C_6-\frac45(3\alpha-2\beta+3)(1+\nu)-\frac85(2\alpha-3\beta+1)\nu\Big)\nonumber\\
&+&\frac{p_r^2}{r}\Big(-4C_1-3C_6-\frac45(3\alpha-2\beta+3)(1+\nu)+\frac85(4\alpha-2\beta+5)\nu\Big)\nonumber\\
&+&\frac{1}{r^2}\Big(-2C_6+\frac85(3\alpha-2\beta+3)(\nu-3)+\frac8{15}(12\alpha-9\beta+11)\nu\Big)\Big]
\,.
\eea

\end{widetext}


\begin{thebibliography}{99}


%\cite{Blanchet:2013haa}
\bibitem{Blanchet:2013haa}
L.~Blanchet,
``Post-Newtonian Theory for Gravitational Waves,''
%Living Rev. Rel. \textbf{17}, 2 (2014);
%doi:10.12942/lrr-2014-2\\
 Living Rev.Rel. \textbf{27}, 1, 4 (2024)
[arXiv:1310.1528 [gr-qc]].


%\cite{Damour:1981bh}
\bibitem{Damour:1981bh}
T.~Damour and N.~Deruelle,
``Radiation Reaction and Angular Momentum Loss in Small Angle Gravitational Scattering,''
Phys. Lett. A \textbf{87}, 81 (1981)
%doi:10.1016/0375-9601(81)90567-3


\bibitem{D1982} 
T.~Damour;
``Probl\`eme des deux corps et freinage de rayonnement en relativit\'e
g\'en\'erale,''
 C.R.\ Acad.\ Sc.\ Paris, S\'erie II, {\bf 294}, pp 1355-1357 (1982)


%\cite{Damour:1983tz}
\bibitem{Damour:1983tz}
T.~Damour,
``Gravitational radiation reaction in the binary pulsar and the quadrupole formula controversy,''
Phys. Rev. Lett. \textbf{51}, 1019-1021 (1983)
%doi:10.1103/PhysRevLett.51.1019



%\cite{Jaranowski:1996nv}
\bibitem{Jaranowski:1996nv}
P.~Jaranowski and G.~Schaefer,
``Radiative 3.5 post-Newtonian ADM Hamiltonian for many body point - mass systems,''
Phys. Rev. D \textbf{55}, 4712-4722 (1997)
%doi:10.1103/PhysRevD.55.4712


%\cite{Pati:2002ux}
\bibitem{Pati:2002ux}
M.~E.~Pati and C.~M.~Will,
``Post-Newtonian gravitational radiation and equations of motion via direct integration of the relaxed Einstein equations. 2. Two-body equations of motion to second post-Newtonian order, and radiation reaction to 3.5 post-Newtonian order,''
Phys. Rev. D \textbf{65}, 104008 (2002)
%doi:10.1103/PhysRevD.65.104008
[arXiv:gr-qc/0201001 [gr-qc]].


%\cite{Nissanke:2004er}
\bibitem{Nissanke:2004er}
S.~Nissanke and L.~Blanchet,
``Gravitational radiation reaction in the equations of motion of compact binaries to 3.5 post-Newtonian order,''
Class. Quant. Grav. \textbf{22}, 1007-1032 (2005)
%doi:10.1088/0264-9381/22/6/008
[arXiv:gr-qc/0412018 [gr-qc]].


%\cite{Blanchet:1996vx}
\bibitem{Blanchet:1996vx}
L.~Blanchet,
``Gravitational radiation reaction and balance equations to post-Newtonian order,''
Phys. Rev. D \textbf{55}, 714-732 (1997)
%doi:10.1103/PhysRevD.55.714
[arXiv:gr-qc/9609049 [gr-qc]].


%\cite{Blanchet:2024loi}
\bibitem{Blanchet:2024loi}
L.~Blanchet, G.~Faye and D.~Trestini,
``Gravitational radiation reaction for compact binary systems at the fourth-and-a-half post-Newtonian order,''
Class. Quant. Grav. \textbf{42}, no.6, 065015 (2025)
%doi:10.1088/1361-6382/adac9d
[arXiv:2407.18295 [gr-qc]].


%\cite{Blanchet:2026suq}
\bibitem{Blanchet:2026suq}
L.~Blanchet, G.~Faye, E.~Seraille and D.~Trestini,
``Gravitational radiation reaction for compact binary systems at the fourth-and-a-half post-Newtonian order in harmonic coordinates,''
Class. Quant. Grav. \textbf{43}, no.10, 105009 (2026)
%doi:10.1088/1361-6382/ae6411
[arXiv:2601.06743 [gr-qc]].



%\cite{Galley:2012qs}
\bibitem{Galley:2012qs}
C.~R.~Galley and A.~K.~Leibovich,
``Radiation reaction at 3.5 post-Newtonian order in effective field theory,''
Phys. Rev. D \textbf{86}, 044029 (2012)
%doi:10.1103/PhysRevD.86.044029
[arXiv:1205.3842 [gr-qc]].


%\cite{Leibovich:2023xpg}
\bibitem{Leibovich:2023xpg}
A.~K.~Leibovich, B.~A.~Pardo and Z.~Yang,
``Radiation reaction for nonspinning bodies at 4.5PN in the effective field theory approach,''
Phys. Rev. D \textbf{108}, no.2, 024017 (2023)
%doi:10.1103/PhysRevD.108.024017
[arXiv:2302.11016 [gr-qc]].



%\cite{Iyer:1995rn}
\bibitem{Iyer:1995rn} 
  B.~R.~Iyer and C.~M.~Will,
  ``Post-Newtonian gravitational radiation reaction for two-body systems: Nonspinning bodies,''
  Phys.\ Rev.\ D {\bf 52}, 6882 (1995).
%doi:10.1103/PhysRevD.52.6882


%\cite{Blanchet:1993ng}
\bibitem{Blanchet:1993ng}
L.~Blanchet,
``Time asymmetric structure of gravitational radiation,''
Phys. Rev. D \textbf{47}, 4392-4420 (1993)
%doi:10.1103/PhysRevD.47.4392


%\cite{Konigsdorffer:2003ue}
\bibitem{Konigsdorffer:2003ue}
C.~Konigsdorffer, G.~Faye and G.~Schaefer,
``The Binary black hole dynamics at the third-and-a-half postNewtonian order in the ADM formalism,''
Phys. Rev. D \textbf{68}, 044004 (2003)
%doi:10.1103/PhysRevD.68.044004
[arXiv:gr-qc/0305048 [gr-qc]].


%\cite{Buonanno:1998gg}
\bibitem{Buonanno:1998gg}
A.~Buonanno and T.~Damour,
``Effective one-body approach to general relativistic two-body dynamics,''
Phys. Rev. D \textbf{59}, 084006 (1999)
%doi:10.1103/PhysRevD.59.084006
[arXiv:gr-qc/9811091 [gr-qc]].


%\cite{Buonanno:2000ef}
\bibitem{Buonanno:2000ef}
A.~Buonanno and T.~Damour,
``Transition from inspiral to plunge in binary black hole coalescences,''
Phys. Rev. D \textbf{62}, 064015 (2000)
%doi:10.1103/PhysRevD.62.064015
[arXiv:gr-qc/0001013 [gr-qc]].


%\cite{Bini:2012ji}
\bibitem{Bini:2012ji}
D.~Bini and T.~Damour,
``Gravitational radiation reaction along general orbits in the effective one-body formalism,''
Phys. Rev. D \textbf{86}, 124012 (2012)
%doi:10.1103/PhysRevD.86.124012
[arXiv:1210.2834 [gr-qc]].


%\cite{Gopakumar:1997ng}
\bibitem{Gopakumar:1997ng}
A.~Gopakumar, B.~R.~Iyer and S.~Iyer,
``Second postNewtonian gravitational radiation reaction for two-body systems: Nonspinning bodies,''
Phys. Rev. D \textbf{55}, 6030-6053 (1997)
[erratum: Phys. Rev. D \textbf{57}, 6562 (1998)]
%doi:10.1103/PhysRevD.57.6562
[arXiv:gr-qc/9703075 [gr-qc]].

%\cite{Khalil:2021txt}
\bibitem{Khalil:2021txt}
M.~Khalil, A.~Buonanno, J.~Steinhoff and J.~Vines,
``Radiation-reaction force and multipolar waveforms for eccentric, spin-aligned binaries in the effective-one-body formalism,''
Phys. Rev. D \textbf{104}, no.2, 024046 (2021)
%doi:10.1103/PhysRevD.104.024046
[arXiv:2104.11705 [gr-qc]].


%\cite{Ramos-Buades:2021adz}
\bibitem{Ramos-Buades:2021adz}
A.~Ramos-Buades, A.~Buonanno, M.~Khalil and S.~Ossokine,
``Effective-one-body multipolar waveforms for eccentric binary black holes with nonprecessing spins,''
Phys. Rev. D \textbf{105}, no.4, 044035 (2022)
%doi:10.1103/PhysRevD.105.044035
[arXiv:2112.06952 [gr-qc]].


%\cite{Gamboa:2024imd}
\bibitem{Gamboa:2024imd}
A.~Gamboa, M.~Khalil and A.~Buonanno,
``Third post-Newtonian dynamics for eccentric orbits and aligned spins in the effective-one-body waveform model seobnrv5ehm,''
Phys. Rev. D \textbf{112}, no.4, 044037 (2025)
%doi:10.1103/rb1c-nx5f
[arXiv:2412.12831 [gr-qc]].


%\cite{Damour:2004bz}
\bibitem{Damour:2004bz}
T.~Damour, A.~Gopakumar and B.~R.~Iyer,
``Phasing of gravitational waves from inspiralling eccentric binaries,''
Phys. Rev. D \textbf{70}, 064028 (2004)
%doi:10.1103/PhysRevD.70.064028
[arXiv:gr-qc/0404128 [gr-qc]].


%\cite{Bini:2025rng}
\bibitem{Bini:2025rng}
D.~Bini, A.~Geralico and S.~Rufrano Aliberti,
``Radiation-reaction correction to scattering binary dynamics at the next-to-leading post-Newtonian order,''
Phys. Rev. D \textbf{112}, no.10, 104005 (2025)
%doi:10.1103/ggbn-fw7j
[arXiv:2509.17853 [gr-qc]].


%\cite{Damour:2000ni}
\bibitem{Damour:2000ni}
T.~Damour, P.~Jaranowski and G.~Schaefer,
``Equivalence between the ADM-Hamiltonian and the harmonic coordinates approaches to the third postNewtonian dynamics of compact binaries,''
Phys. Rev. D \textbf{63}, 044021 (2001)
[erratum: Phys. Rev. D \textbf{66}, 029901 (2002)]
%doi:10.1103/PhysRevD.63.044021
[arXiv:gr-qc/0010040 [gr-qc]].


%\cite{Blanchet:2002mb}
\bibitem{Blanchet:2002mb}
L.~Blanchet and B.~R.~Iyer,
``Third postNewtonian dynamics of compact binaries: Equations of motion in the center-of-mass frame,''
Class. Quant. Grav. \textbf{20}, 755 (2003)
%doi:10.1088/0264-9381/20/4/309
[arXiv:gr-qc/0209089 [gr-qc]].


\end{thebibliography}
\end{document}